\documentclass[fleqn,usenatbib]{mnras}

\usepackage{newtxtext,newtxmath}

\usepackage[T1]{fontenc}

\DeclareRobustCommand{\VAN}[3]{#2}
\let\VANthebibliography\thebibliography
\def\thebibliography{\DeclareRobustCommand{\VAN}[3]{##3}\VANthebibliography}

\usepackage{graphicx}	
\usepackage{amsmath}	
\usepackage{amssymb}	
\usepackage{multirow}
\usepackage{array,booktabs}
\usepackage{nccmath}
\usepackage{gensymb}
\usepackage[normalem]{ulem}
\usepackage[dvipsnames]{xcolor}

\newcommand{\Se}{\texttt{Sedonu }}
\newcommand{\se}{\texttt{Sedonu}}
\newcommand{\Fla}{\texttt{FLASH }}
\newcommand{\fla}{\texttt{FLASH}}

\title[ASL calibration for BNS merger simulations]{Calibration of the Advanced Spectral Leakage scheme for neutron star merger simulations, and extension to smoothed-particle hydrodynamics}

\author[Gizzi D. et al.]{
D. Gizzi,$^{1}$\thanks{E-mail: davide.gizzi@astro.su.se}
C. Lundman,$^{1}$
E. O'Connor,$^{1}$
S. Rosswog,$^{1}$
A. Perego$^{2,3}$
\\
$^{1}$The Oskar Klein Centre, Department of Astronomy, Albanova, Stockholm University, SE-106 91 Stockholm, Sweden\\
$^{2}$Dipartimento di Fisica, Universit\`a degli Studi di Trento, via Sommarive 14, 38123 Trento, Italy\\
$^{3}$INFN-TIFPA, Trento Institute for Fundamental Physics and Applications, via Sommarive 14, I-38123 Trento, Italy
}

\date{Accepted XXX. Received YYY; in original form ZZZ}

\pubyear{2020}

\begin{document}
\label{firstpage}
\pagerange{\pageref{firstpage}--\pageref{lastpage}}
\maketitle

\begin{abstract}
We calibrate a neutrino transport approximation,
called Advanced Spectral Leakage (ASL), 
with the purpose of modeling 
neutrino-driven winds in neutron star mergers. 
Based on a number of snapshots we gauge the ASL parameters by
comparing against both the two-moment (M1) scheme
implemented in the \Fla code
and the Monte Carlo neutrino code \se. 
The ASL scheme contains 
three parameters, the least robust 
of which results to be 
a blocking parameter for 
electron neutrinos and anti-neutrinos. 
The parameter steering
the angular distribution of 
neutrino heating is re-calibrated
compared to the earlier work. We also 
present a new, fast and
mesh-free algorithm for 
calculating spectral optical depths, which, when using 
Smoothed Particle Hydrodynamics (SPH), 
makes the neutrino transport completely particle-based.
We estimate a speed-up
of a factor of $\gtrsim 100$ 
in the optical depth calculation 
when comparing to a grid-based approach.
In the suggested calibration we recover luminosities 
and mean energies within $25\%$.
A comparison of the 
rates of change of internal
energy and electron fraction 
in the neutrino-driven wind suggests 
comparable accuracies of ASL and M1, but 
a higher computational efficiency of the ASL scheme.
We estimate that
the ratio between 
the CPU hours spent on the ASL 
neutrino scheme 
and those spent on the hydrodynamics
is $\lesssim 0.8$ per timestep when
considering the SPH code 
MAGMA2 as source
code for the Lagrangian 
hydrodynamics, to be
compared with a 
factor of 10 from the M1 
in \fla. 
\end{abstract}




\section{Introduction}
The combined detection of
gravitational and electromagnetic
waves from the neutron star merger event GW170817 \citep{Abbott2017a,Abbott2017c} 
has solved several decades-old 
puzzles.
For example, the detection of 
the short Gamma-Ray
Burst (sGRB) GRB170817A $\sim 1.74$ 
s after the merger
\citep{Savchenko2017,Goldstein2017b} 
confirmed the long suspected
association between sGRBs
and compact binary mergers 
\citep{eichler89,Paczynski1986,Narayan1992}. Moreover,
the detection of the macronova/kilonova transient AT2017gfo
\citep{Chornock17,Kasliwal17,Kilpatrick17,Pian17,
Coulter2017,Soares2017,Lipunov2017,Arcavi2017,
Evans2017,Smartt17} confirmed binary neutron star
mergers as a major, and possibly dominant,
r-process nucleosynthesis site \citep{Rosswog18a,Kasen2017,Drout2017},
thus confirming earlier theoretical predictions on the subject 
\citep{lattimer74,eichler89,rosswog99,freiburghaus99b}.
The event GW170817 also allowed to place
important constraints
on the Equation of State (EoS) of nuclear matter \citep{Most2018,Soumi2018,Abbott2018,Radice2018,
Radice2019,Coughlin2019,Kiuchi2019,Bauswein2017,Jiang2019,Jiang2020} and provided an independent
measurement of the Hubble parameter \citep{Abbott2017b,Dhawan2020}.\\
The luminous 
blue component of the macronova 
transient AT2017gfo has also 
emphasized the role played by
neutrinos in changing the electron 
fraction $Y_e$ of the ejected matter.
Weak interactions  modify
the electron fraction in a fair portion of the
ejecta from initial values of $Y_e\sim 0.05$
to values above $Y_e=0.25$, 
where the resulting nucleosynthesis
changes abruptly \citep{korobkin12a,lippuner15} 
and no more lanthanides 
and heavier nuclei are produced.
This increase in $Y_e$ also drastically 
reduces the optical
opacities \citep{kasen13a,tanaka13a} 
and leads to blue (rather than red)
transients after about a day rather than a week.  
Fits to the blue component of the light curve from macronova models
suggest ejecta of several $10^{-2}\:M_{\odot}$
with velocity $\sim 0.3c$.
There is a general agreement that secular mass ejection
from the remnant disk is needed to achieve such large ejecta
amounts
(e.g. \cite{Ciolfi2020,Nedora2020,Radice2018b}),
but it is still a matter of debate
to which extent the different proposed 
secular ejection channels contribute to the blue
macronova.
The ejection channels include winds driven by
magnetic fields
\citep{Ciolfi2017,Ciolfi2020}, viscosity
\citep{Fernandez2013,
Fernandez2019,Fujibayashi2018,
Fujibayashi2020,Siegel2017}, neutrinos
\citep{Dessart2009,Perego2014,Martin2015} 
and spiral-waves
\citep{Nedora2019,Nedora2020}.
Our motivation here are neutrino-driven winds, 
for which no systematic study  in dynamical
simulations of binary neutron star mergers
exists to date. For this reason,
we have recently implemented 
an extension to the original Advanced
Spectral Leakage (ASL) scheme \citep{Perego2016}
with the purpose of modelling the
neutrino absorption responsible
for launching the winds in merger 
simulations \citep{Gizzi2019}. 

In this paper we extend the 
analysis of the ASL scheme
by performing a careful 
calibration for the 
three parameters that 
enter the scheme.
These parameters
have an impact on 
both neutrino emission and absorption,
and must therefore be gauged carefully.
Two of these parameters were already 
introduced in \cite{Perego2016}, but
calibrated for core-collapse supernovae
simulations. 
They define the neutrino emission and 
represent two physical effects. 
The first is a blocking parameter, accounting for both Pauli blocking effects due to the 
fermionic nature of neutrinos and 
for inward neutrino fluxes when
calculating the neutrino emission above the
decoupling region, while 
the second is a thermalization parameter,
describing
the number of weak 
interactions needed 
to thermalize 
neutrinos.
We added a third 
parameter in \cite{Gizzi2019} specifically
for modelling the neutrino absorption
responsible for launching the 
neutrino-driven winds, and
we gauged it by comparing the outcome
of the simulations with the two-moment scheme
(M1) approach implemented in \Fla 
\citep{Fryxell2000,OConnor2015,Oconnor2018}. 
Here, we use both the M1
and the Monte Carlo neutrino transport
code \Se \citep{Richers2015}
as sources of comparison.
We use neutrino quantities directly
impacted by each parameter to perform our
calibration and we extract them by
post-processing
a number of snapshots of 
binary neutron star remnants. 

When using the ASL scheme, we 
present a new, mesh-free
implementation for calculating spectral,
multi-group optical depths for
Smoothed-Particle Hydrodynamics (SPH) 
\citep{monaghan92,Monaghan05,Rosswog2009,Rosswog15,Rosswog2015_2,Rosswog2020}
simulations. This makes our ASL 
fully mesh-free
and ideal for 
dynamical SPH simulations of 
merging neutron stars
with neutrino effects. 

The paper is structured as follows.
We begin with a description of the
neutrino transport codes
in Sec.~\ref{sec:nuapproaches}.
We then describe a new particle-based
algorithm to calculate the optical depth
in Sec.~\ref{sec:pbasedopt},
where we also include a comparison with grid-based
calculations.
Sec.~\ref{sec:calibration} describes 
the set of snapshots
that we use, as well as the calibration
strategy.
The results of the calibration 
are discussed in Sec.~\ref{sec:results}.
In Sec.~\ref{sec:conclusions}
we summarise and draw our conclusions.

\section{Neutrino transport methods}
\label{sec:nuapproaches}
The neutrino transport approaches 
we use for our analysis are:
the Monte Carlo code \se,
the two-moment scheme (M1) developed in
the Eulerian hydrodynamics code \fla, 
and the ASL scheme.
All of them are spectral,
(i.e., neutrino energy is treated 
as an independent variable 
and neutrino-matter 
interactions are calculated 
for different
neutrino energies),
an important
ingredient given the 
energy-squared 
dependence entering neutrino
cross sections 
(see e.g. \cite{Burrows2004} 
and references therein). Moreover, 
we choose the same set of weak interactions:
production and absorption
of electron neutrinos and anti-neutrinos 
via charged current processes involving 
nucleons and nuclei, neutrino emission
by bremsstrahlung and pair processes,
and finally elastic
scattering off nucleons and nuclei.
We choose the SFHo EoS \citep{Steiner2013} to 
calculate spectral emissivities and opacities.
\Se and M1 interpolate them from 
the \texttt{NuLib} tables \citep{OConnor2015}.
The ASL scheme calculates them
following \cite{Bruenn1985,Mezzacappa1999,Hannestad1998}.
Three distinct neutrino 
species are modelled: 
electron neutrinos $\nu_e$, 
electron anti-neutrinos $\bar{\nu}_e$ and
a collective species $\nu_x$ 
for heavy-lepton neutrinos, 
comprising $\mu$ and $\tau$ neutrinos
and anti-neutrinos.

\subsection{Monte Carlo}
\label{sec:MC}

Monte Carlo methods can be used 
to solve the Boltzmann equation. 
They provide an exact solution to 
the transport problem in the limit 
of an infinite particle number. 
However, Monte Carlo neutrino 
transport methods tend to be 
computationally expensive, since
a large number of particles is 
needed to keep the numerical noise
in the solution at a low level. 
These methods have therefore mostly 
been used on static, spherically 
symmetric core-collapse supernovae 
snapshots \citep{Janka1989, Janka1989b, Janka1992, Keil2003, Abdikamalov2012}. 
Similarly, studies of Monte Carlo 
neutrino transport in neutron star mergers have 
been carried out by post-processing 
snapshots, assuming the neutrino 
field can be approximated as 
steady-state \citep{Richers2015}.
Recently, \citet{Foucart2020} 
performed the first dynamical
hydrodynamic simulation of merging 
neutron stars up to $\sim 5$ ms 
post-merger with a computationally 
particularly efficient Monte Carlo approach. 
Here, we use the steady-state Monte Carlo code 
\Se
\citep{Richers2015} on axisymmetric 
fluid snapshots to calibrate the parameters
of our ASL scheme.

We use a logarithmically spaced 
energy grid that spans the range 
$E_{\nu} \in [0.5,200]$ MeV.
 Neutrinos are simulated as packets
 passing through and interacting 
 with a static fluid background 
 imported from hydrodynamic 
 simulations.
 Each packet is specified
 by the neutrino energy $E_{\nu}$, 
 the location \textbf{x}, the unit 
 vector $\hat{\textbf{d}}$ describing
 the propagation direction, and the
 total number of neutrinos in the packet. 
 A fixed number of neutrino packets 
 are emitted from each grid cell. 
 In particular, \Se randomly samples
 the energy-dependent emissivity 
 to determine the neutrino energy,
 and the packet direction of propagation
 is drawn from an isotropic distribution. 
 Neutrino packets interact with the 
 fluid by being partially absorbed 
 during the packet propagation, 
 and changing propagation direction
 when scattered. The distance 
 the packet travels before 
 interacting is randomly drawn
 from an exponential distribution
 that depends on the scattering 
 mean free path. If the packet is
 scattered, it is given an isotropically
 random new direction. While moving,
 the packet deposits both energy 
 and lepton number
 to the traversed  
 grid-cells, from which the 
 corresponding rates of change 
 can be derived.  
 We emit $100$ Monte Carlo packets 
 per energy group and grid cell. 
 In this way, integrated quantities
 such as luminosities and mean neutrino
 energies are stable in spite of 
 Monte Carlo noise, which is estimated
 to be a thousand times smaller.

\subsection{Two-moment scheme (M1)}
\label{sec:M1}
A moment scheme is an approximation to the
Boltzmann neutrino transport, obtained by 
evaluating angular integrals of the Boltzmann equation in order to time-evolve the 
angular moments of the distribution function
\citep{Lindquist1966,Bruenn1978,Bruenn1985,Mezzacappa1999}.
The M1 scheme is a 
multi-dimensional, spectral
approach where only the 
0th and the 1st moments of the neutrino 
distribution function are evolved,
which respectively
describe the spectral energy density and 
the energy flux density \citep{castor2004}. 
The system of equations is closed by 
an assumed closure relation, 
which is often analytical
\citep{Audit2002,Levermore1981,
Minerbo1978,Smith1997,Pons2000}.
Generally, the analytical 
form captures
the physics of the transport well
in specific regimes, 
while approximating the 
exact solution in others. 
Neutrino-matter interactions are included
in appropriate source terms that appear on the
right-hand side of the moment equations
and which are solved via 
techniques that are borrowed from
finite volume hydrodynamics.

The M1 scheme has been widely applied
in simulations
that model shock revival in 
core-collapse supernovae 
\citep{OConnor2015, Oconnor2018, OConnor2013, Obergaulinger2014, Skinner2019}
and recently compared to other transport 
approaches \citep{pan19,Cabezon2018,OConnor2018b},
showing good performance and agreement
with alternative schemes at the level of
$10-20\%$ in luminosities, mean energies
and shock radii. However, not so much
has been done in the context 
of binary neutron star mergers. 
In the work of \cite{Foucart2016a}
an M1 scheme has been
adopted to test the impact of the 
energy spectrum on the composition of
merger outflows. In a subsequent work
\cite{Foucart2018} compared a grey
M1 scheme against a Monte Carlo 
approach, showing the limitations 
of the M1 analytical closure in 
recovering the properties of the 
polar ejecta and therefore in
modelling neutrino-driven winds 
and kilonovae.
Another quantitative study 
of the assumptions
employed in analytical closures
and their violation
has been carried out more recently
by post-processing 
snapshots with 
Monte Carlo transport \citep{Richers2020}.
Despite their limitations,
\cite{Foucart2020} have shown that
time-dependent neutrino transport in 
an M1 scheme with an analytical closure that carefully
treats the energy spectrum approximates the 
Monte Carlo solution  within $\sim 10-20\%$ in the
composition and velocity of ejecta, 
as well as electron neutrino and
anti-neutrino luminosities and mean
energies. 
In Sec.~\ref{sec:results} we 
explore the performance of the M1
scheme implemented in \Fla and compare it 
with our new ASL treatment and the Monte Carlo code \se.

\subsection{Advanced Spectral Leakage (ASL)}
\label{sec:ASL}
We start by briefly summarizing
the original implementation of the
ASL scheme in its main aspects.
For more details, the reader is referred to  \cite{Perego2014,Perego2016}. Subsequently, we 
describe the implementation of
\cite{Gizzi2019}.

The ASL is a  
spectral neutrino scheme, 
ideal for computationally inexpensive, 
multi-dimensional hydrodynamic simulations 
of core-collapse supernovae \citep{Perego2016,pan19,Ebinger2020,Curtis2019,Ebinger2020_2} and binary neutron star mergers 
\citep{Perego2014} with neutrino transport effects. 
We assume a logarithmically-spaced
energy grid 
spanning the range [3,300] MeV. 
The change in the internal energy 
and electron fraction of the matter
due to weak interactions is estimated 
by means of calculating \textit{local}
neutrino emission and absorption rates.
However, their computation depends 
on an integrated quantity, the spectral, 
flavour-dependent \textit{optical depth}, 
defined at position \textbf{x} as:

\begin{ceqn}
\begin{equation}
    \label{eq:opdep}
    \tau_{\nu}(E,\textbf{x})= \int_{\gamma:\textbf{x}\rightarrow+\infty}
    \frac{1}{\lambda_{\nu}(E,\textbf{x'}(s))}{\rm d}s,
\end{equation}
\end{ceqn}
where $\lambda_{\nu}(E,\textbf{x'})$
is the neutrino mean free path for
neutrinos of species $\nu$ at energy
$E$ and position \textbf{x'}, 
calculated over a path $\gamma$. 
Here we introduce a new, 
mesh-free algorithm to compute 
$\tau_{\nu}(E,\textbf{x})$ 
in a SPH context, so
that neither the hydrodynamics nor 
the  neutrino transport
depend on any grid implementation,
see Sec.~\ref{sec:pbasedopt} 
for a detailed description.
Depending on the interactions 
that enter the computation of 
$\lambda_{\nu}(E,\textbf{x'})$, we can 
distinguish between the \textit{total} 
optical depth $\tau_{\nu,\rm{tot}}(E,\textbf{x})$, 
accounting equally for both absorption 
and elastic scattering interactions, 
and the \textit{energy} optical depth $\tau_{\nu,\rm{en}}(E,\textbf{x})$, 
giving more emphasis to energy-exchange 
interactions between neutrinos and matter.
From the above definitions of 
$\tau_{\nu}$ different regimes can be defined:\\\\
$\bullet$ The \textit{equilibrium-diffusive} 
regime, where $\tau_{\nu,\rm{tot}} \gg 1$ and $\tau_{\nu,\rm{en}} \gtrsim 1$. 
Neutrinos can be treated as a
trapped Fermi gas in thermal 
and weak equilibrium with the matter.\\\\
$\bullet$ The \textit{diffusive} regime, 
where $\tau_{\nu,\rm{tot}} \gg 1$ and $\tau_{\nu,\rm{en}} \lesssim 1$. 
Neutrinos still diffuse, but
they are not necessarily in 
thermal equilibrium with the matter.\\\\
$\bullet$ The \textit{semi-transparent} 
regime, where $\tau_{\nu,\rm{tot}} \sim 1$. 
Although neutrinos can still be 
partially absorbed, they begin to 
decouple from matter, and the surfaces 
around which this occur are called neutrino-surfaces, which can be identified when
$\tau_{\nu} \sim 2/3$. 
Low-energy neutrinos decouple earlier
than high-energy neutrinos, 
and therefore the neutrino surfaces 
for the latters are wider
\citep{Endrizzi2020}. 
In the context of binary neutron
star mergers, neutrino absorption
occurring in this regime 
(hereafter referred to as \textit{heating})
drives mass ejection via  
so-called \textit{neutrino-driven winds}
\citep{Perego2014,Dessart2009}.\\\\
$\bullet$ In the \textit{free-streaming regime},
where $\tau_{\nu,\rm{tot}} \lesssim 1$, locally produced neutrinos stream 
out freely, and neutrino 
absorption is negligible.\\\\
The local emission rate is given by
\begin{equation}
\label{eq:ser}
    r_{\nu}(E,\textbf{x})= (1 - \alpha_{\nu,\mathrm{blk}})\Tilde{r}_{\nu}(E,\textbf{x})\frac{1}{\Psi_{\nu}(\textbf{x})}\mathrm{exp}(-\tau_{\nu,\mathrm{en}}(E,\textbf{x})/\tau_{\mathrm{cut}}),
\end{equation}
where $\Tilde{r}_{\nu}(E,\textbf{x})$ is given as smooth interpolation between the production rate $r_{\rm{\nu,prod}}(E,\textbf{x})$
and the diffusion rate $r_{\rm{\nu,diff}}(E,\textbf{x})$:
\begin{ceqn}
\begin{equation}
\label{eq:emrate}
    \Tilde{r}_{\nu}(E,\textbf{x})= \frac{r_{\mathrm{\nu,prod}}(E,\textbf{x})\: r_{\mathrm{\nu,diff}}(E,\textbf{x})}{r_{\mathrm{\nu,prod}}(E,\textbf{x})+r_{\mathrm{\nu,diff}}(E,\textbf{x})},
\end{equation}
\end{ceqn}
and 
\begin{ceqn}
\begin{equation}
    \Psi_{\nu}(\textbf{x})= \frac{\int_0^{+\infty}\Tilde{r}_{\nu}(E,\textbf{x})e^{-\tau_{\nu,\mathrm{en}}(E,\textbf{x})/\tau_{\mathrm{cut}}}E^2{\rm d}E}{\int_0^{+\infty}\Tilde{r}_{\nu}(E,\textbf{x})E^2{\rm d}E}.
\end{equation}
\end{ceqn}
Eq.~\ref{eq:ser} depends 
on two parameters, so far 
calibrated in the context 
of core-collapse supernovae simulations:\\\\
1) a blocking parameter 
$\alpha_{\nu,\rm{blk}}$, 
that accounts 
for Pauli blocking
effects due to the
large amount of neutrinos 
locally produced or emitted 
at the neutrino surface, and 
for the fact
that the interpolation given by
Eq.~\ref{eq:emrate} favours the
production rate in the free-streaming
regime. Neutrino emission in
this regime is assumed to be isotropic,
and a fraction of neutrinos is 
therefore emitted inward and does
not contribute to the luminosity.\\\\
2) a thermalization parameter
$\tau_{\rm{cut}}$, representing 
the amount of interactions 
needed to thermalize neutrinos, 
and which accounts for
neutrinos that exchange energy 
with the fluid while propagating
from the innermost to the 
outermost regions of the remnant,
making the neutrino spectrum 
at the decoupling surface softer.\\\\
The computation of the heating rate
requires the knowledge of the 
distribution of the neutrino 
fluxes coming from the region
inside the neutrino surfaces.
While the original implementation
of the ASL assumes spherically 
symmetric neutrino fluxes 
\citep{Perego2016}, the more 
complex geometry of a binary
neutron star merger causes 
anisotropic neutrino fluxes \citep{Perego2014,Dessart2009,Foucart2016a,Rosswog2003}.
We have recently 
implemented an extension of 
the original ASL scheme
that allows the modelling of
neutrino-driven winds 
\citep{Gizzi2019}, which we briefly recap here.\\
The presence of an opaque disk 
around the central remnant of 
a binary neutron star merger 
makes it difficult for neutrinos
to escape along the equatorial region. 
Therefore an observer located far
from the decoupling region is 
expected to receive the maximum 
neutrino flux at angles near 
the poles, and the minimum flux
along the equatorial region itself \citep{Rosswog2003}. 
Accordingly, we expect most 
of the neutrino heating to occur
along the polar regions.
We abbreviate as
$n_{\nu,\tau \lesssim 1}(E,\textbf{x})$ 
the number of neutrinos with 
energy between $E$ and $E+dE$ available 
for absorption at position 
\textbf{x} in the semi-transparent 
and free streaming regimes.
It can be approximated as:

\begin{ceqn}
\begin{equation}
    n_{\mathrm{\nu,\tau \lesssim 1}}(E,\textbf{x})= \frac{(1+p)(1+\beta_{\nu}\cos^p(\theta))}{1+p+\beta_{\nu}} \frac{l_{\mathrm{\nu}}(E,|\textbf{x}|)}{4\pi |\textbf{x}|^2c\mu_{\mathrm{\nu}}(E,\textbf{x})},
    \label{eq:new_nudensity}
\end{equation}
\end{ceqn}
where the first term 
depends of the polar angle
$\theta$ defined with respect 
to the polar axis of the
remnant, 
and the second term 
is similar to the one in \cite{Perego2016}, 
with a spherically symmetric 
part divided by the 
flux factor  $\mu_{\nu}(E,\textbf{x})$.
The angular term is assumed 
to be axially-symmetric 
around the polar axis, 
since some increasing degree 
of axial symmetry 
is generally expected from a 
few tens of ms irrespective 
of the mass ratio \citep{Perego2019}. 
Moreover, it contains the quantity 
$\beta_{\nu}$, which is defined as:

\begin{ceqn}
\begin{equation}
\label{eq:beta}
\beta_{\nu}= \frac{\Lambda_{\nu}(\theta \approx 0 \degree)}
{\Lambda_{\nu}(\theta \approx 90 \degree)} - 1,
\end{equation}
\end{ceqn}
$\Lambda_{\nu}(\theta)$ being the 
luminosity per solid angle at 
$\theta$.
The computation of 
$\Lambda_{\nu}(\theta)$ is 
done with a spectral version 
of the prescription of 
\cite{Rosswog2003}. 
In the context of merger simulations
$\beta_{\nu} > 0$, while 
for spherically symmetric 
fluxes $\beta_{\nu} = 0$ 
and we recover the neutrino 
density form of \cite{Perego2016}. 
Eq.~(\ref{eq:new_nudensity}) 
introduces a third parameter, 
the heating parameter $p$, 
providing the flux modulation 
as a function of $\theta$. 
The spherically symmetric term 
contains the spectral number of 
neutrinos $l_{\nu}$ per unit time 
of energy $E$ at radius $R=|\textbf{x}|$,
calculated by following \citep{Gizzi2019}. 
The flux factor is computed 
analytically, similarly to \cite{Oconnor2010}:
\begin{equation}
\label{eq:finalfluxf}
\frac{1}{\mu_{\mathrm{\nu}}}(E,\textbf{x})= \begin{cases} 1.5\:\tau_{\mathrm{\nu,tot}}(E,\textbf{x}) + 1 &\text{if  $\tau_{\mathrm{\nu,tot}}(E,\textbf{x}) \leq 2/3$}\\
2 &\text{$\mathrm{otherwise}$}
\end{cases}.
\end{equation}
The flux factor describes the average cosine of the propagation angle of the streaming neutrinos. For an observer far from the neutrino surfaces the neutrino distribution function peaks radially, therefore $\mu_{\nu} \rightarrow 1$. On the other hand, for an observer close to the neutrino surface $\mu_{\nu} \rightarrow 1/2$ \citep{Liebendorfer2009}. \\
The parameter $p$ has been previously calibrated \citep{Gizzi2019} by comparing the heating rate maps of electron neutrinos and anti-neutrinos\footnote{Re-absorption by heavy-lepton neutrinos
is negligible, see \cite{Gizzi2019} for details.}
for a binary neutron star merger snapshot against those obtained with the M1 scheme in \Fla \citep{Fryxell2000,Oconnor2018}. This demonstrated 
that the new ASL implementation  is  able to capture the heating rate distribution to better than a factor of 2, with the largest deviations right above the central remnant. In this work we further improve the analysis of the new ASL scheme and we provide  values for $\alpha_{\nu,\rm{blk}}$, $\tau_{\rm{cut}}$ and $p$ that are more accurate for  neutron star mergers. All the details are described in Sec.
~\ref{sec:calibration}. 

\section{Particle-based optical depth}
\label{sec:pbasedopt}

\begin{figure}
\centering
\includegraphics[width = 1 \linewidth]{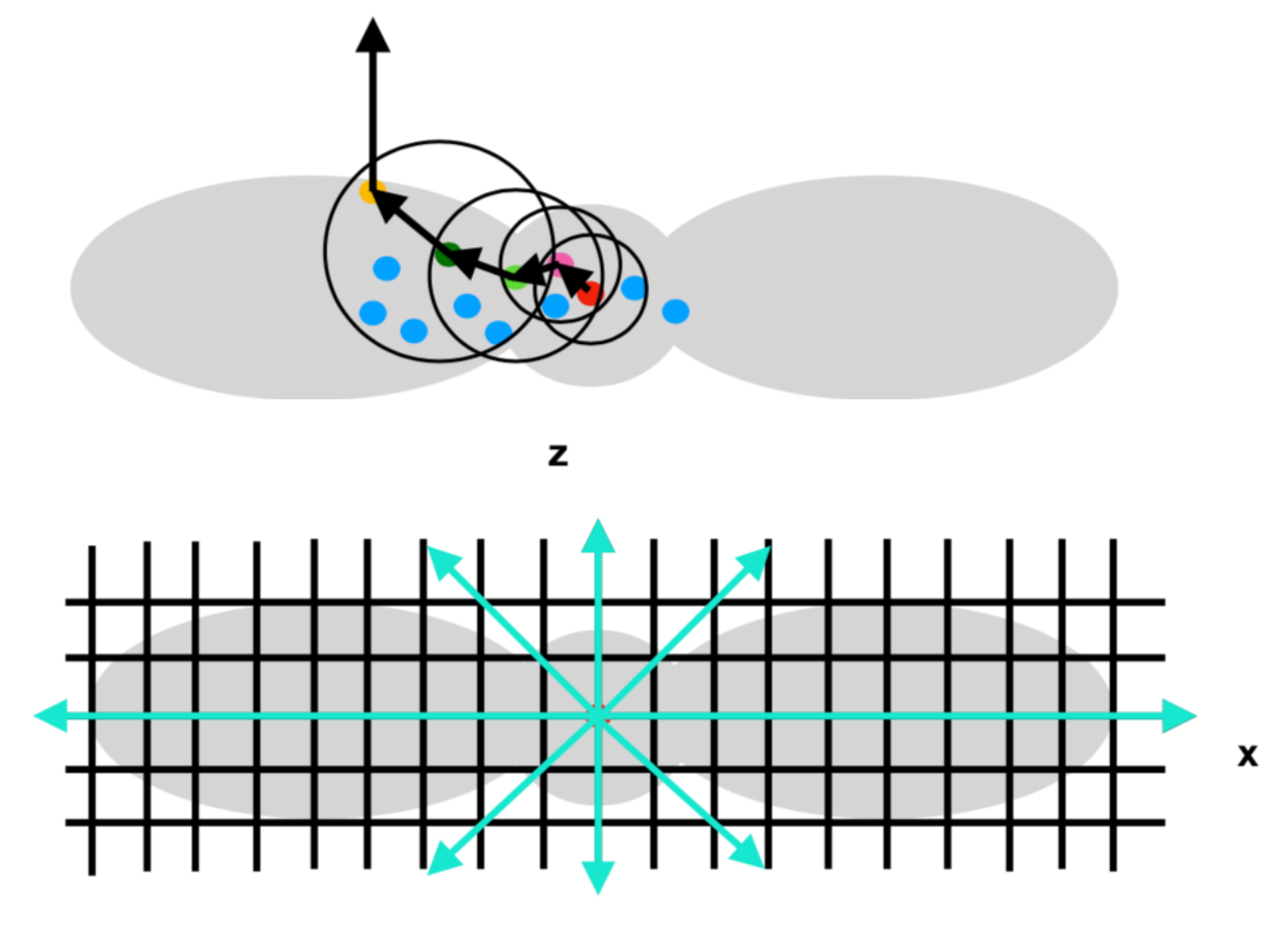}
\caption{Simplified sketches of 
the particle-based (top) and 
grid-based (bottom) approaches
for the optical depth 
calculation. Shown in
grey is a neutron star remnant surrounded
by a disk. In the particle-based
model, each particle has a smoothing length
defining the radius of each circle. 
The latter encloses the particle neighbours.
Unlike the grid-based approach, where
outgoing radial paths are a priori selected for
calculating the optical depth, 
in the particle-based case 
the path along 
which neutrinos move is defined
by the density distribution, 
and therefore it can be non-straight.}
\label{fig:sketchtau}
\end{figure}

\begin{figure*}
\begin{minipage}{0.7
\linewidth}
\centering
\includegraphics[width = 1 \linewidth]{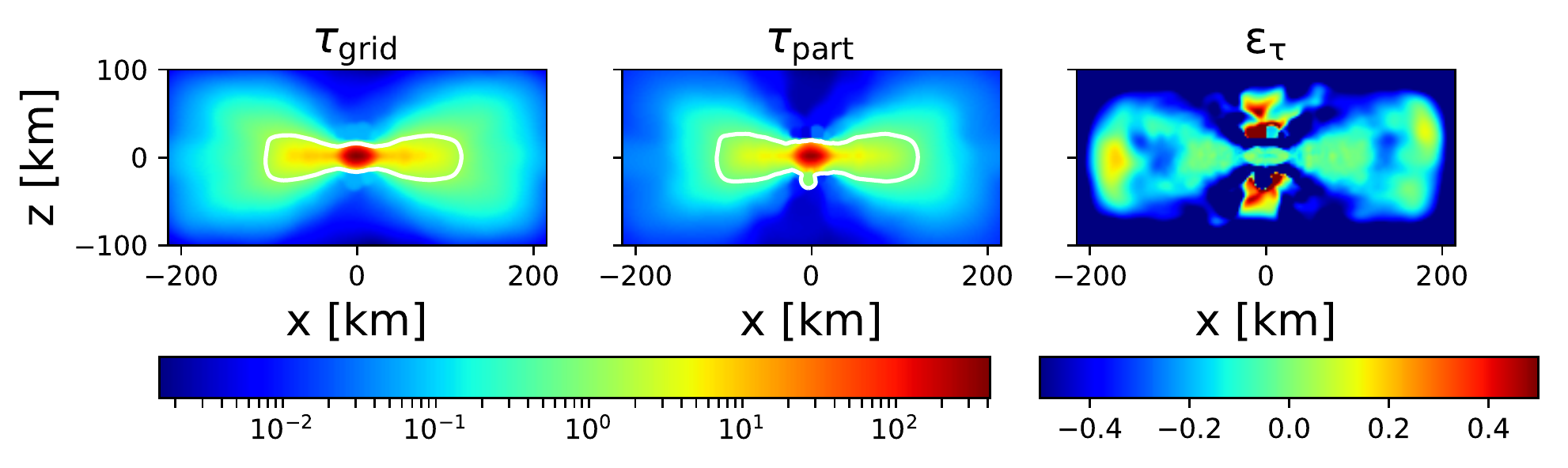}
\end{minipage}\\
\begin{minipage}{0.7 \linewidth}
\centering
\includegraphics[width = 1 \linewidth]{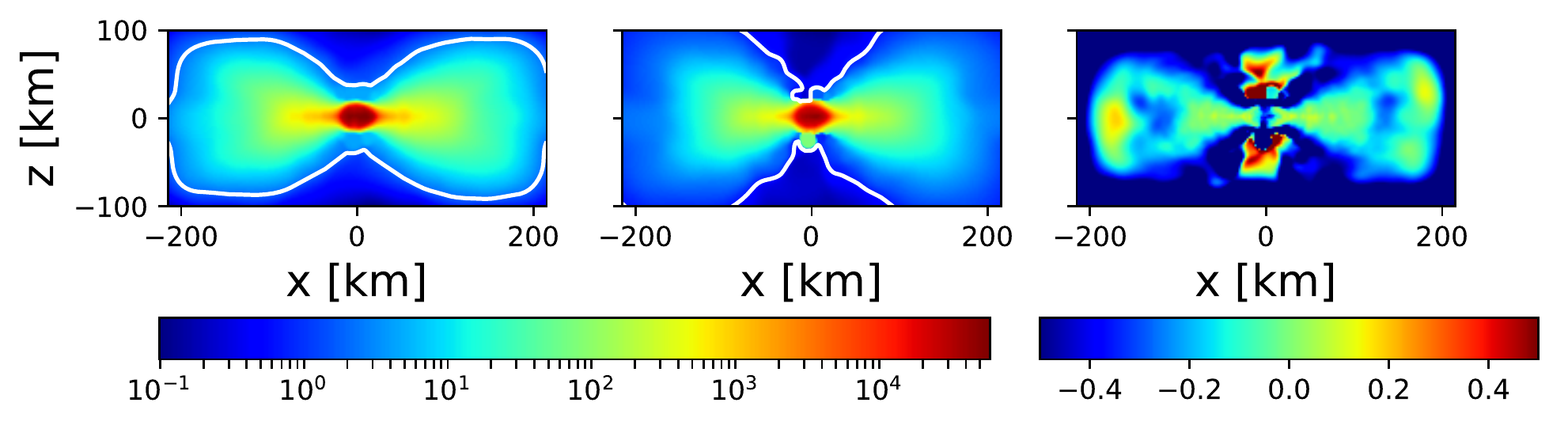}
\end{minipage}\\
\vspace{1.1cm}
\begin{minipage}{0.7 \linewidth}
\centering
\includegraphics[width = 1 \linewidth]{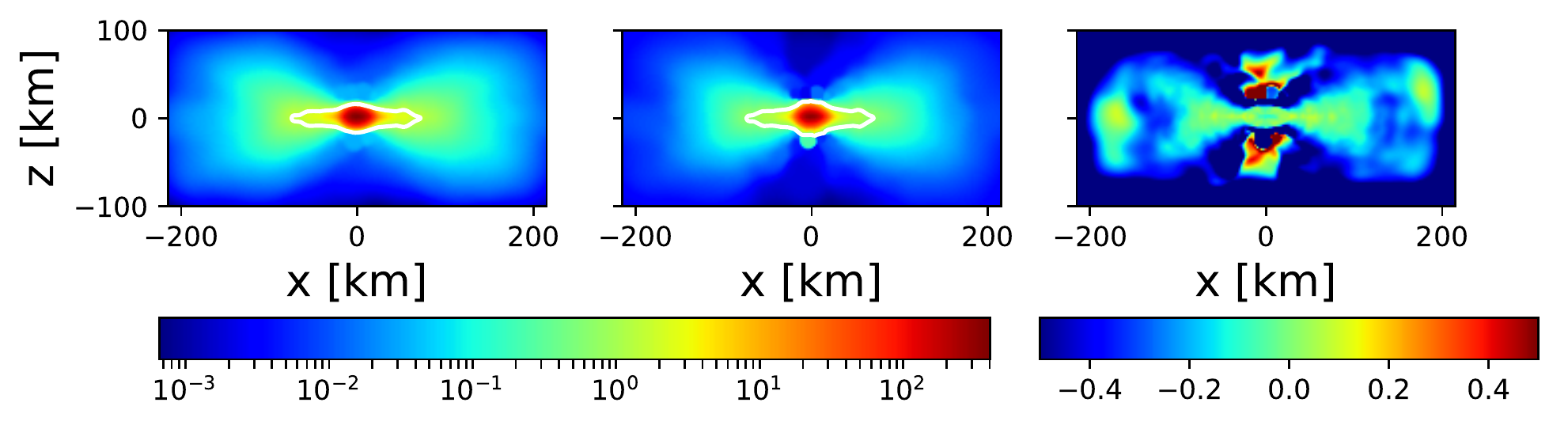}
\end{minipage}\\
\begin{minipage}{0.7 \linewidth}
\centering
\includegraphics[width = 1 \linewidth]{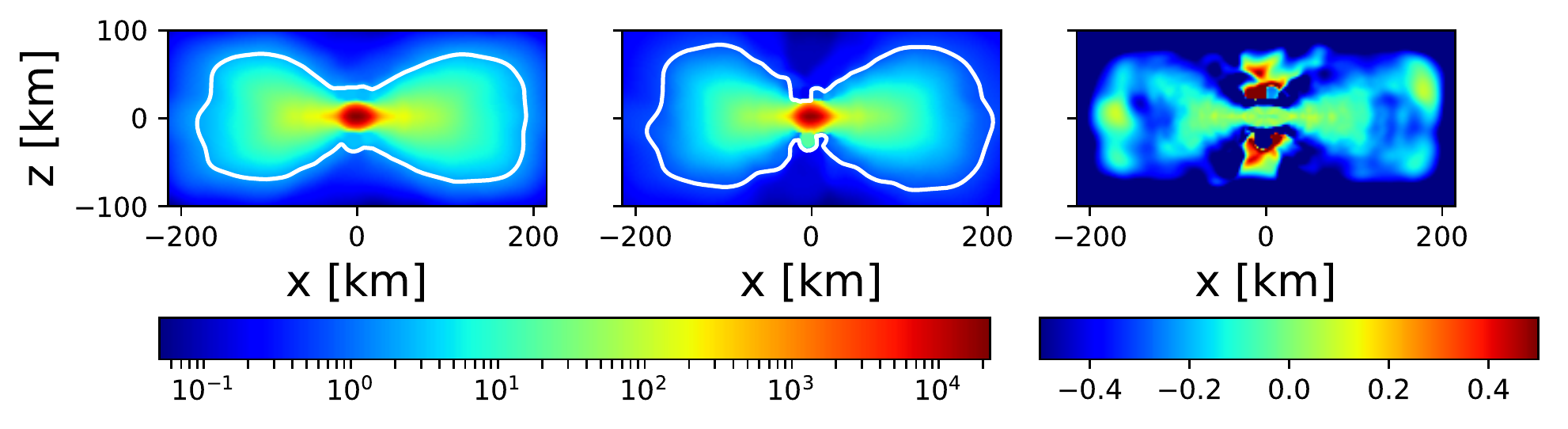}
\end{minipage}\\
\vspace{1.1cm}
\begin{minipage}{0.7 \linewidth}
\centering
\includegraphics[width = 1 \linewidth]{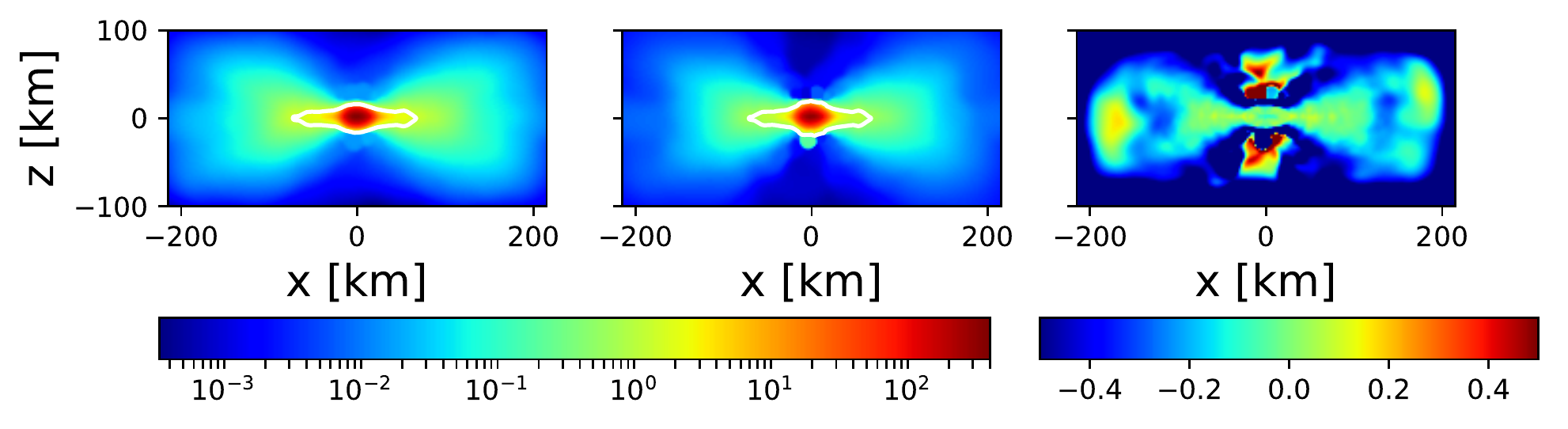}
\end{minipage}\\
\begin{minipage}{0.7 \linewidth}
\centering
\includegraphics[width = 1 \linewidth]{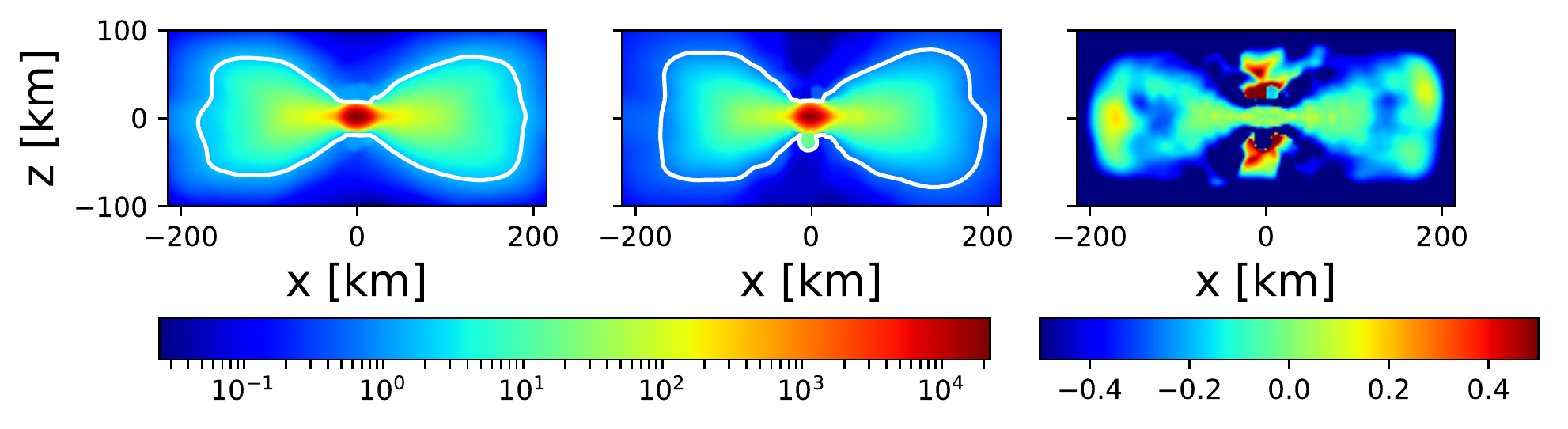}
\end{minipage}
\caption{Total optical depth on the x-z plane for electron neutrinos (top), electron anti-neutrinos (middle) and heavy-lepton neutrinos (bottom). For each species we show the map for both a low  energy ($\sim 10$ MeV, top) and a high energy ($\sim 100$ MeV, bottom) bin of the spectrum. Moreover, from left to right, we show the maps obtained from the grid-based and particle-based calculations, and a map of the relative difference $\epsilon_{\tau}= \frac{\tau_{\rm{grid}}-\tau_{\rm{part}}}{\tau_{\rm{grid}}}$, respectively.}
\label{fig:tautot}
\end{figure*}

\begin{figure*}
\begin{minipage}{0.7
\linewidth}
\centering
\includegraphics[width = 1 \linewidth]{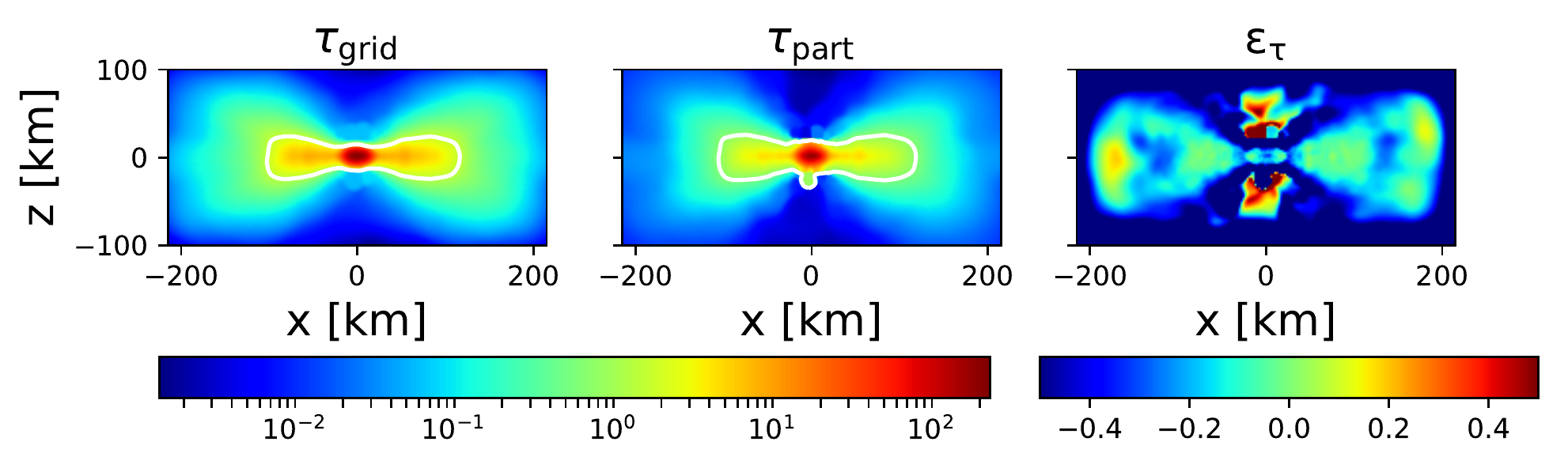}
\end{minipage}\\
\begin{minipage}{0.7 \linewidth}
\centering
\includegraphics[width = 1 \linewidth]{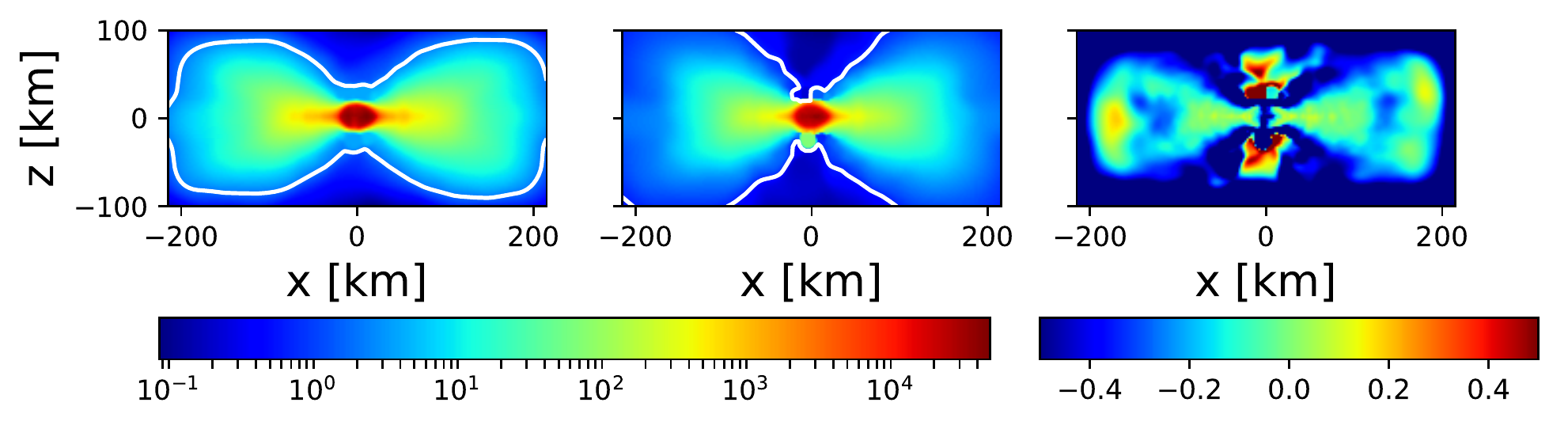}
\end{minipage}\\
\vspace{1.1cm}
\begin{minipage}{0.7 \linewidth}
\centering
\includegraphics[width = 1 \linewidth]{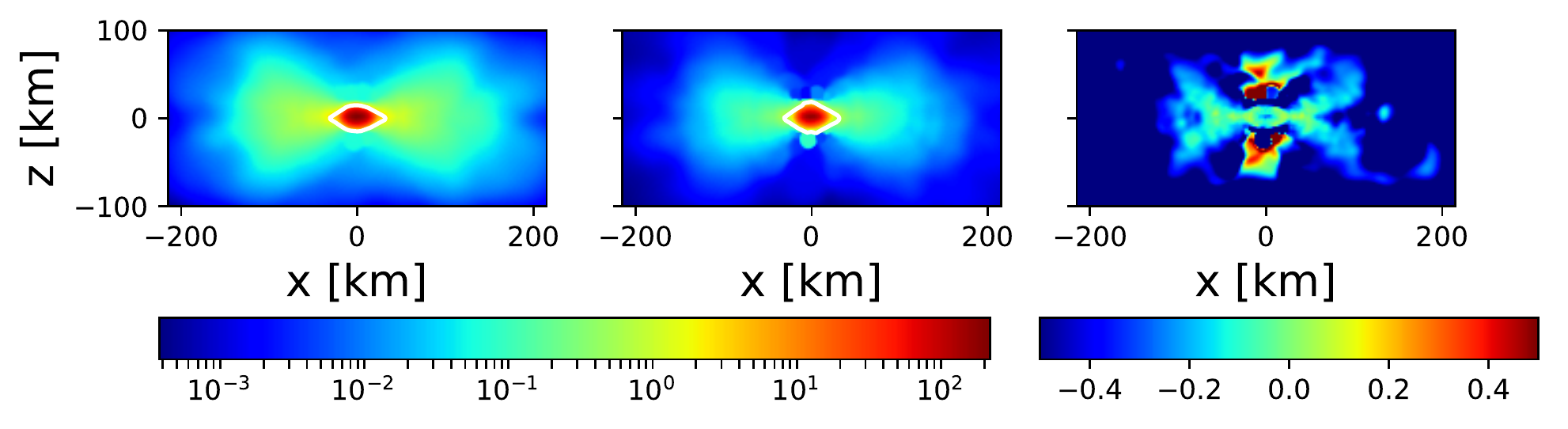}
\end{minipage}\\
\begin{minipage}{0.7 \linewidth}
\centering
\includegraphics[width = 1 \linewidth]{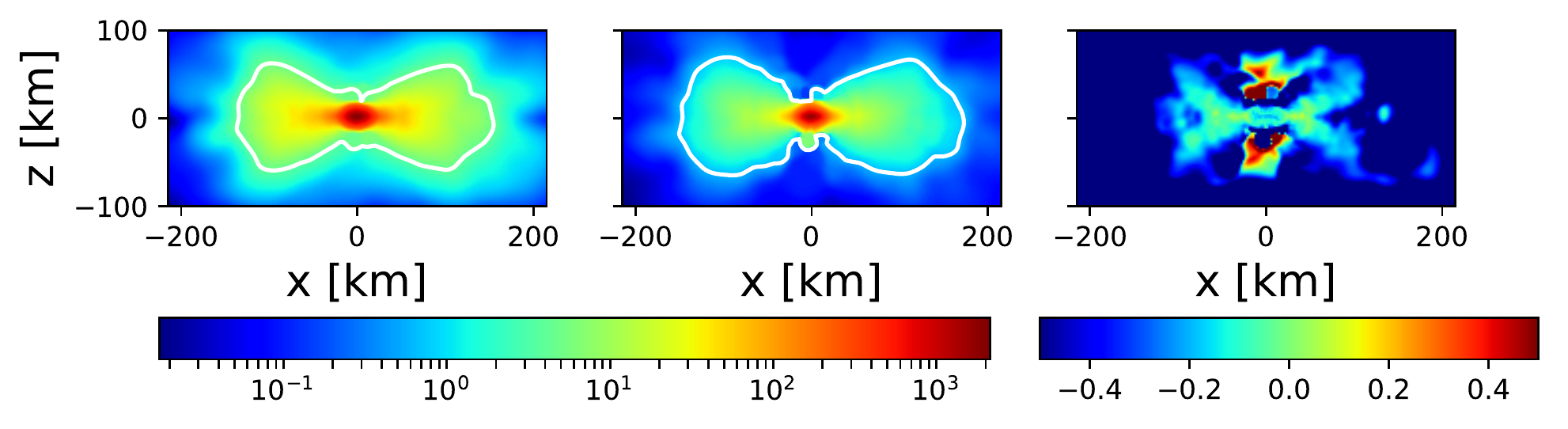}
\end{minipage}\\
\vspace{1.1cm}
\begin{minipage}{0.7 \linewidth}
\centering
\includegraphics[width = 1 \linewidth]{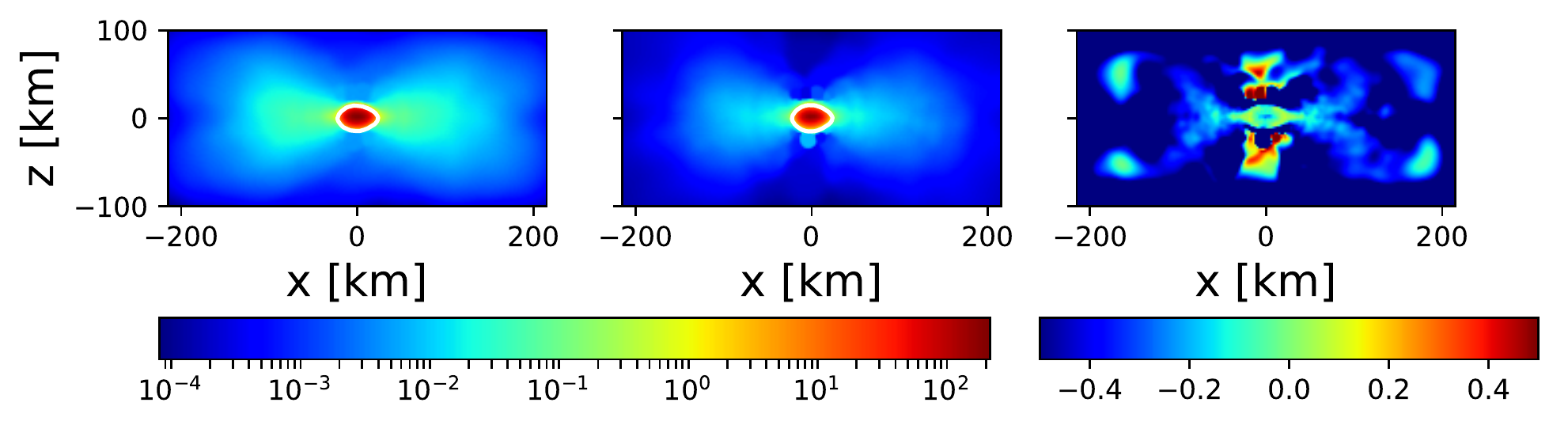}
\end{minipage}\\
\begin{minipage}{0.7 \linewidth}
\centering
\includegraphics[width = 1 \linewidth]{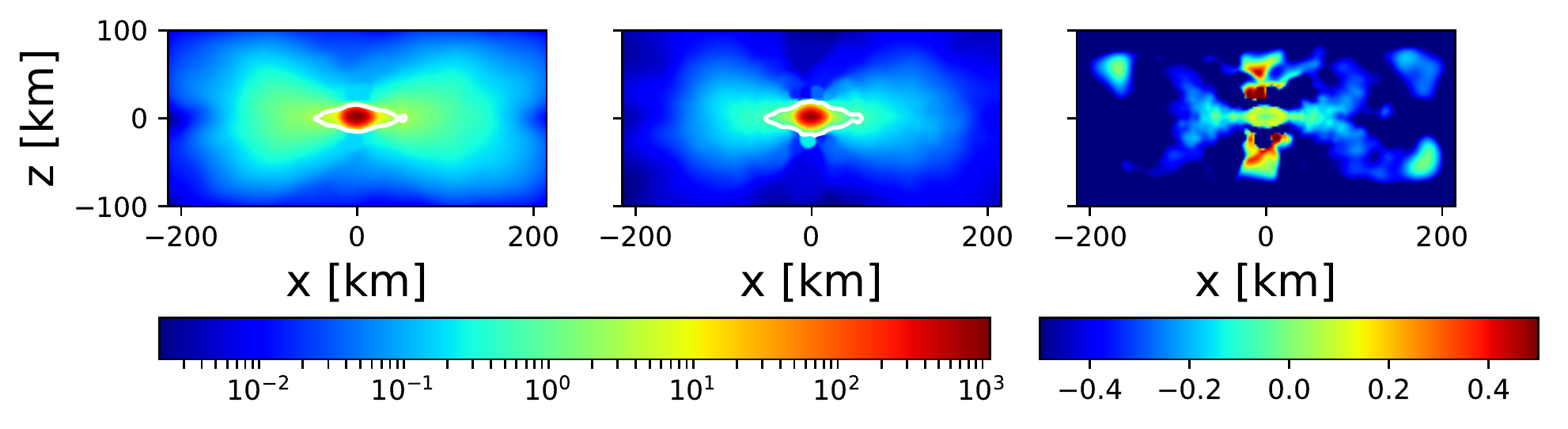}
\end{minipage}
\caption{Same as Fig.~\ref{fig:tautot}, but for the energy optical depth.}
\label{fig:tauen}
\end{figure*}

SPH is a mesh-free, particle-based method
to solve the ideal hydrodynamics equations, see  \cite{Monaghan05,Rosswog2009,Rosswog15,Rosswog2015_2,Rosswog2020} for reviews of
and recent developments within this method.\\
Each particle $a$ has a \textit{smoothing length} $h_a$, which defines the particle's
local interaction radius. In particular, given two particles 
\textit{a} and \textit{b}, 
the latter is a neighbour of 
the former if the distance 
$\Delta r_{ab}= |\textbf{x}_a - \textbf{x}_b|$
between \textit{a}
and \textit{b} satisfies 
$\Delta r_{ab} \leq 2h_a$.
From Eq.~\ref{eq:opdep} we can
define the optical depth at 
$\textbf{x}_a$ as:

\begin{ceqn}
\begin{equation}
    \label{eq:opdep2}
    \tau_{\nu}(E,\textbf{x}_a)= \int_{\gamma:\textbf{x}_a\rightarrow+\infty}
    \frac{1}{\lambda_{\nu}(E,\textbf{x'}(s))}{\rm d}s,
\end{equation}
\end{ceqn}
or equivalently:
\begin{ceqn}
\begin{equation}
    \tau_{\nu}(E,\textbf{x}_a)= \int_{\gamma:\textbf{x}_a\rightarrow+\infty}
    \kappa(E,\textbf{x'}(s))\rho(\textbf{x'}(s)){\rm d}s,
    \label{eq:tau_A2}
\end{equation}
\end{ceqn}
where $1/\lambda_{\nu}(E,\textbf{x'}(s))= \kappa(E,\textbf{x'}(s))\rho(\textbf{x'}(s))$, 
with $\kappa(E,\textbf{x'})$ being
the opacity for energy $E$ at position \textbf{x'} 
and $\rho(\textbf{x'})$ the density at 
position \textbf{x'}. Obviously, the optical depth
at a given location depends on the exact path $\gamma$
along which the neutrino escapes. Neutrinos that are produced 
inside a neutron star merger remnant will typically
escape from the remnant after a potentially large number 
of interactions. While each radiation particle follows 
an individual path, particles have larger probabilities
to escape if they are locally moving in a direction of 
increasing mean free path. We try to mimick 
this behaviour in our meshfree algorithm to 
efficiently calculate optical depths. 
We approximate the integral in Eq.~(\ref{eq:tau_A2})
as a sum of contributions between pairs of SPH-particles 
so that we move from SPH-particle to SPH-particle
until a transparent region has been reached. 
The choice of the next SPH particle on the route to escape,
is guided by the aim to pick the direction that maximizes 
the local mean free path. In practice we use the mass
density as a proxy, i.e. each SPH particle chooses as 
the next particle the one in its neighbour list that 
has the smallest density. In other words we are  
following the local negative density gradient until 
neutrinos can escape freely.\\
More specifically, we proceed via the following steps: 
\begin{enumerate}
    \item The SPH particle for which we search the optical depth
    is labelled $a$. For the convenience of the subsequent notation
    we also identify particle $a$ as the "zeroth neighbour particle" and label it $b_0$.
    \item The first task is to find the particle $b_1$ in $a$'s
    neighbour list that has the minimum density
    among the neighbours and add the contribution, $\tau_{b_0 \rightarrow b_1}$ to the optical depth of $a$.
    \item Once $b_1$ has been found, find particle $b_2$ in $b_1$'s neighbour list 
    that has the minimum density and so on until a particle $b_{k_\infty}$ has been found from which neutrinos can escape freely.
We use as escape condition  $\rho \lesssim 10^9 \mathrm{g~cm^{-3}}$ \citep{Endrizzi2020}.
\end{enumerate}
In other words, we  discretize the integral of Eq.~\ref{eq:opdep2} as:
\begin{equation}
    \tau_{\nu,a}(E)= \sum_{k=0}^{k_{\rm esc}}\tau_{b_k \rightarrow b_{k+1}}(E),
\end{equation}
where
\begin{equation}
\begin{split}
\tau_{b_k \rightarrow b_{k+1}}(E)= & \int_{\vec{x}_{b_k}}^{\vec{x}_{b_{k+1}}}\frac{ds}{\lambda_{\nu}(E,\textbf{x'}(s))}\\ \approx &
\frac{|\vec{x}_{b_{k+1}} - \vec{x}_{b_{k}} |}{2} \left(\frac{1}{\lambda_{b_{k+1}}(E)} + \frac{1}{\lambda_{b_{k}}(E)}\right)
\end{split}
\end{equation}
and we have applied the trapezoidal rule in the last step.
The local mean free paths at the particle positions are calculated following equations 1 and 2 of \cite{Perego2016},
for the total and the energy mean free paths respectively.
\\
Unlike in \cite{Gizzi2019} and
in most of the literature
(see \cite{Perego2014,George2020,Endrizzi2020} as examples), 
where the  
standard approach is to
pre-select 
radial paths on a grid,
calculate the integral of
Eq.~\ref{eq:opdep} over 
these paths, and then take
the minimum optical depth, 
here there is no assumption
on the type of path, but it
is the matter itself 
that tells neutrinos how
to leak out
(see also \cite{Perego2014b}
for a similar
methodology). 
Fig.~\ref{fig:sketchtau} shows
sketches of both the particle- (top) 
and grid-based (bottom) methods. 
The main advantage
of our new approach is the 
affordable computational 
cost required to get
spectral optical
depths compared to 
grid-based integration methods. 
This is on one side due 
to the fact that we use
directly the particle
properties for the 
calculations, and on the
other side because 
we adopt the fast recursive
coordinate bisection 
tree of \cite{Gafton2011}
for neighbour search. \\
Figs.~\ref{fig:tautot}-\ref{fig:tauen}
show on the 
plane x-z the 
total and energy optical depths
respectively, for case 1) 
described in Sec.~\ref{sec:setup}.
Both of them are calculated 
on a grid with the
algorithm of \cite{Gizzi2019},
and with our particle-based
approach. We SPH-map on 
the plane x-z the particle-based
optical depths for comparison purposes.
We show the case for 
a low-energy 
bin ($\sim 10$ MeV, upper panels) and for a 
high-energy bin ($\sim 100$ MeV, lower panels)
of our energy grid, 
and for electron neutrinos (top), 
electron anti-neutrinos (middle),
and heavy-lepton neutrinos
(bottom). We also show
the map of the relative difference $\epsilon_{\tau}= (\tau_{\rm{grid}}-\tau_{\rm{part}})/\tau_{\rm{grid}}$.
The optical depth appears to be
larger along the 
disk with respect to
the grid-based approach, 
and lower 
along the poles.
Low density regions in SPH
are not as well resolved as 
the high density ones, 
therefore the SPH maps 
in the former regime are only 
indicative of what the optical
depth could actually be
(although we expect it to be
rather small anyway). 
Note that $\tau_{\rm{grid}}$
is calculated once initial 
conditions of 
density, temperature and electron
fraction of the 
SPH snapshot are mapped on 
the grid. Therefore,
$\tau_{\rm{grid}}$
suffers from the same
uncertainties of 
$\tau_{\rm{part}}$ 
in low density regions.
The higher optical depths
along the disk with 
the particle-based approach
leads to more extended 
neutrino surfaces
at high neutrino energies,
especially for electron
neutrinos, which are 
the most interacting species
given the neutron richness 
of the material.
Nevertheless, our algorithm well
captures the expected distribution
of the neutrino surfaces 
according to different neutrino
energies and species 
(see \cite{Gizzi2019,Perego2014} 
for details). Accounting for 
different resolutions that can be
used both in
the particle- and grid-based approaches, 
we estimate the particle-based
algorithm to be $\gtrsim 100$ 
times faster than the grid-based
one.

\section{Calibration}
\label{sec:calibration}
\subsection{Setup}
\label{sec:setup}
\begin{figure*}
\centering
\includegraphics[width = 1 \linewidth]{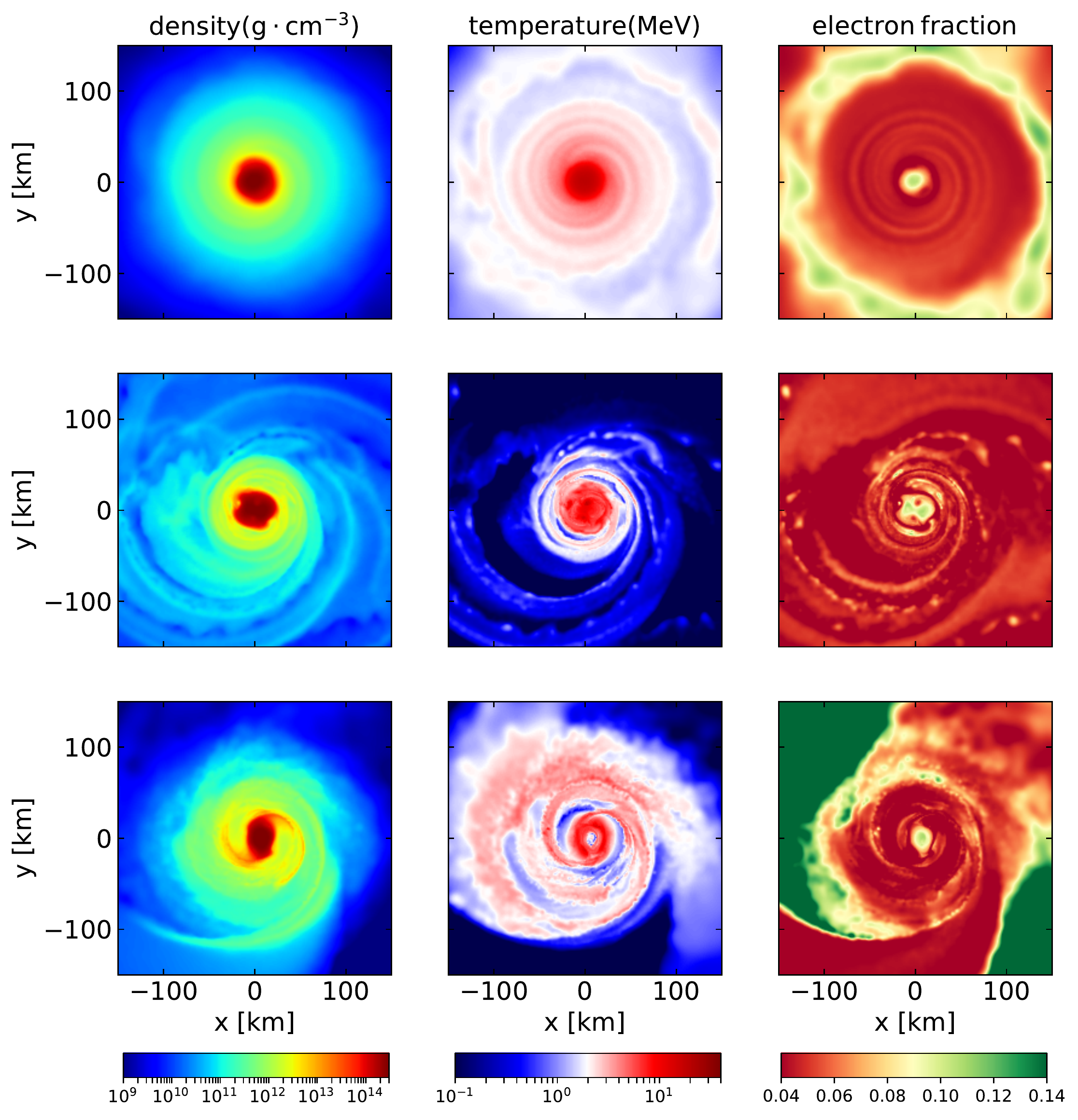}
\caption{Density (left column), 
temperature (middle column),
and electron 
fraction (right column) maps on the 
equatorial plane at equilibrium, 
for 
a 1.4-1.4 $M_{\odot}$ binary 
neutron star merger at 
$\approx 38$ ms after merger
(top row), 
a 1.3-1.3 $M_{\odot}$ 
binary neutron star merger
at $\approx 18$ ms after merger
(middle row), and 
a 1.2-1.3 $M_{\odot}$ binary
neutron star merger at 
$\approx 18$ ms after merger
(bottom row).
From top to bottom the 
degree of axial symmetry of
the remnant decreases.
Snapshots are taken from the 
dynamical simulations of \protect\cite{Rosswog17a}.}
\label{fig:xy}
\end{figure*}

\begin{figure*}
\centering
\includegraphics[width = 1 \linewidth]{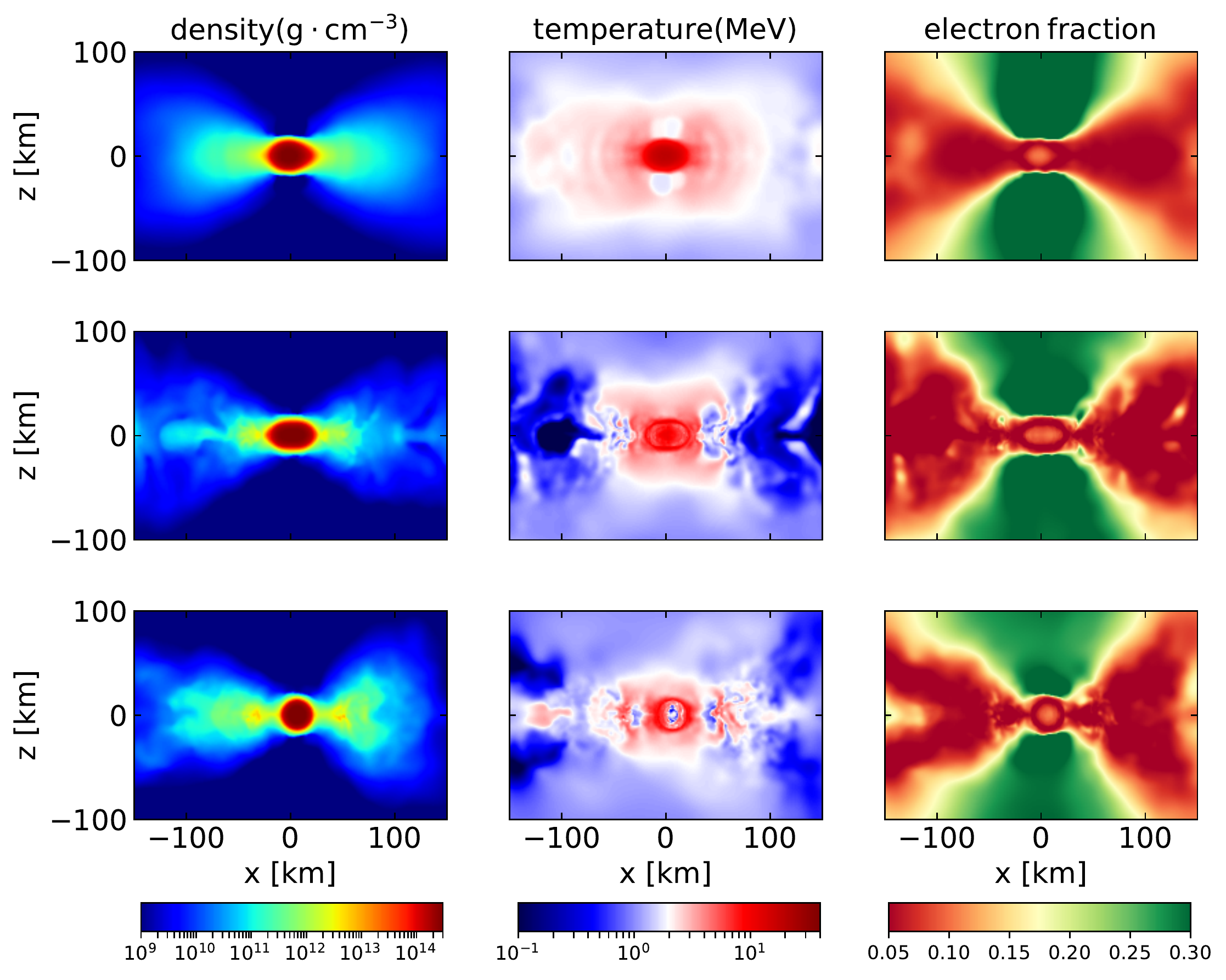}
\caption{Same as in Fig.~\ref{fig:xy},
but on the plane
orthogonal to the
equatorial one.}
\label{fig:xz}
\end{figure*}

We take three snapshots 
of binary neutron star merger remnants from 
the SPH simulations of \cite{Rosswog17a}, 
each one representing a specific
case for calibrating our scheme:\\\\
case 1) an equal mass binary system of 1.4-1.4 $M_{\odot}$ at $\approx 38$ ms after merger. 
This is our late-time configuration, when 
the matter around the remnant has the highest
degree of axial symmetry.\\\\
case 2) an equal mass binary system of 1.3-1.3 $M_{\odot}$ at $\approx 18$ ms after merger. 
In this way we include equal mass 
configurations
at earlier post-merger times, when 
the matter distribution is not
fully axially symmetric.\\\\
case 3) an unequal mass binary system of 1.2-1.3 $M_{\odot}$ at $\approx 18$ ms after merger. 
We consider this case as representative
of unequal mass binaries at early times, 
with the lowest degree of axial symmetry.\\\\
We use cases 2) and 3) to test
the assumption of axial symmetry entering
Eq.~\ref{eq:new_nudensity}.
Table \ref{Table1} summarises 
the properties of each binary configuration.
\begin{table}
\begin{center}
\caption{Summary of the binary properties for each case examined. We report the masses $m_1$ and $m_2$ of each star in the binary, the mass ratio $q=m_2/m_1$ and the time after merger. }
    \begin{tabular}{c|c|c|c|c}
      \hline
      Case & $m_1(M_{\odot})$ & $m_2(M_{\odot})$ & $q$ & \rm{time\:after\:merger\:(ms)}\\ \hline 
      1 & 1.4 & 1.4 & 1.00 & $\approx$ 38\\ \hline
      2 & 1.3 & 1.3 & 1.00 & $\approx$ 18\\ \hline
      3 & 1.3 & 1.2 & 0.92 & $\approx$ 18\\ \hline
    \end{tabular}
    \label{Table1}
    \end{center}
\end{table}
Since the snapshots
are taken 
from simulations
that use a grey leakage scheme
\citep{Rosswog2003},
we first map the SPH properties
on a 3D grid and
let the radiation
field equilibrate with M1 for 3 ms
without changing either the
electron fraction 
or the temperature.
Afterwards, we let 
the electron fraction
evolve for 10 ms with M1 to 
remove grey leakage effects.
We do not consider the temperature
evolution in M1 since we do not
see important differences
between the 
corresponding evolved 
and non-evolved
profiles.
The evolution with M1
allows to
make a more consistent 
comparison between 
neutrino approaches for our
parameter calibration.
Figs.~\ref{fig:xy}-\ref{fig:xz} show
density, temperature and 
electron fraction in
the equatorial plane and on 
the plane orthogonal to it 
after the 10 ms of evolution,
and for each case. 
In order to 
considerably reduce computational costs
when running the Monte Carlo code \Se 
we map density, temperature 
and the evolved electron fraction 
from M1 onto a 2D, axially symmetric 
grid, and run the transport over 
the obtained profiles.
At last, we map the evolved M1 data
back to the SPH particles by tri-linear
interpolation to run the ASL scheme.

\subsection{Strategy}
As a first step we identify  physical quantities 
that are directly impacted by each ASL parameter:\\\\
1) the total luminosities and
mean energies, primarily
affected by $\alpha_{\nu,\rm{blk}}$ 
and $\tau_{\rm{cut}}$.\\\\
2) the neutrino flux at
different polar angles that
an observer located far outside the 
neutrino surfaces would measure.
In this respect, we focus 
on the \textit{trend} of the flux with
the polar angle for 
constraining the parameter $p$ in Eq.~\ref{eq:new_nudensity}.\\\\
Given that the Monte Carlo
approach converges 
(in the limit of infinite particle numbers)
to an exact solution
of the transport equation, 
the best strategy for 
our parameter calibration
would be to extract both quantities 1) and 2) 
from \Se and compare with the ASL scheme.
However, the assumption of
axial symmetry may not be fully appropriate in
all cases, especially
shortly after merger.
We then first test 
the assumption of azimuthal symmetry by 
running the M1 transport
on both 2D and 3D grid setups, and 
compare the neutrino emission maps 
for each species.
For those snapshots where the 
impact on the emission 
is large in at least one of the 
species, we decide to 
consider M1
in 3D for the calibration.
However, given the limitations of
M1 in accurately modelling the
distribution of the 
neutrino flux at different
polar angles \citep{Foucart2018},
we make use of M1 purely for calibrating
$\alpha_{\nu,\rm{blk}}$ 
and $\tau_{\rm{cut}}$, 
while we keep \Se
to calibrate the heating 
parameter $p$. As we shall see,
the 2D assumption does not
impact the trend of the flux 
at different polar angles, but only
its local values. 
On the other hand, 
if the 2D assumption 
does not  
affect the emission of 
any neutrino species
sensitively, 
the calibration of $\alpha_{\nu,\rm{blk}}$,
$\tau_{\rm{cut}}$ and of the heating 
parameter $p$ is performed entirely with \se.\\
The parameter space we explore is 
given by $\alpha_{\nu,\rm{blk}} 
\in [0.35,0.45,0.55,0.65,0.75,0.85]$,
$\tau_{\rm{cut}} 
\in [5,10,15,20]$, and
$p \in [2,4,6,8,10,12]$.
When calibrating, we assume
$\alpha_{\rm{blk}} \equiv 
\alpha_{\nu_e,\rm{blk}}= 
\alpha_{\bar{\nu}_e,\rm{blk}}$, and 
$\alpha_{\nu_x,\rm{blk}}= 0$, similarly 
to \cite{Perego2016}.
We will discuss the 
accuracy of the 
assumption 
$ \alpha_{\nu_e,\rm{blk}}= 
\alpha_{\bar{\nu}_e,\rm{blk}}$
in Sec.~\ref{sec:combination}.
Given that heavy-lepton neutrinos
do not contribute to the heating
either, we decide to neglect
this species entirely 
when calibrating, 
and just focus on the
other two species. We will anyway 
report all values of luminosities and
mean energies for completeness
(see Table \ref{Table2}).\\
During the calibration process, 
when exploring luminosities 
and mean energies,
we first compute:

\begin{ceqn}
\begin{eqnarray}
    \epsilon_{L} &= & 
    \epsilon_{L_{\nu_e}} + \epsilon_{L_{\bar{\nu}_e}}, \\
    \epsilon_{\langle E \rangle} & = & 
    \epsilon_{\langle E_{\nu_e} \rangle} + \epsilon_{\langle E_{\bar{\nu}_e} \rangle},    
\end{eqnarray}
\end{ceqn}

\noindent where 
\begin{ceqn}
\begin{equation}
\epsilon_{L_{i}}= \frac{|L_{i,\rm{ref}}-L_{i,\rm{ASL}}|}{L_{i,\rm{ref}}}
\end{equation}
\end{ceqn}
and 
\begin{ceqn}
\begin{equation}
\epsilon_{\langle E_{i} \rangle}= \frac{|\langle E_{i,\rm{ref}} \rangle - \langle E_{i,\rm{ASL}} \rangle|}{\langle E_{i,\rm{ref}} \rangle},
\end{equation}
\end{ceqn}
$i$ labeling neutrinos of species $i
\in [\nu_e, \bar{\nu}_e]$, while 
$L_{i,\rm{ref}}$ and 
$\langle E_{i,\rm{ref}} \rangle$ are
the luminosity and mean energy of species
$i$ from the reference solution
(either \Se or M1).
We compare
$\epsilon_L$ and 
$\epsilon_{\langle E \rangle}$
and eventually consider 
the error among the two 
more sensitive to a
variation of the parameters.
We look for regions 
in the parameter space 
where this error is minimal
for a first pre-selection,
and subsequently 
explore $\epsilon_{L_i}$ and
$\epsilon_{\langle E_i \rangle}$ 
individually for a more
detailed analysis.\\
When calibrating $p$ with \Se
we assume an analytical form of the  
neutrino fluxes as function 
of the polar angle
in the ASL scheme.
In particular, the flux of 
neutrinos of species $i$ leaving the source
at some distance $R$ is:
\begin{ceqn}
\begin{equation}
    \textbf{F}_{i}= \frac{1}{R^2}\frac{dL_{i}}{d\Omega}\textbf{n},
\end{equation}
\end{ceqn}
\textbf{n} being the unit vector orthogonal 
to the surface $dS= R^2 d\Omega$, and
$d\Omega= d\mathrm{cos}(\theta) d\phi= d\mu d\phi$ 
being the solid angle of polar angle $\theta$
and azimuthal angle $\phi$.
For axial symmetry with
$d\Omega= 2\pi d\mu$ we have: 
\begin{ceqn}
\begin{equation}
    \textbf{F}_{i}= \frac{1}{2\pi R^2}\Big(\frac{dL_{i}}{d\mu}\Big)\textbf{n}.
\end{equation}
\end{ceqn}
Given the analytical angular term in
Eq.~\ref{eq:new_nudensity}, we assume that:
\begin{ceqn}
\begin{equation}
\label{eq:flux1}
    \frac{dL_{i}}{d\Omega}= 
    \frac{1}{4\pi}\frac{(1+p)(1+\beta_{i}\cos^p(\theta))}{1+p+\beta_{i}}L_{i}
\end{equation}
\end{ceqn}
and \begin{ceqn}
\begin{equation}
\label{eq:flux2}
    \frac{dL_{i}}{d\mu}= 
    \frac{1}{2}\frac{(1+p)(1+\beta_{i}\cos^p(\theta))}{1+p+\beta_{i}}L_{i} \, .
\end{equation}
\end{ceqn}
We caution the reader 
that Eq.~\ref{eq:flux2}
is purely to 
understand which value of $p$ 
best reproduces the 
\textit{trend}
of the flux as function of the 
polar angle from \se.
We do not attempt to estimate
the magnitude of the flux at 
each polar angle via Eq.~\ref{eq:flux2}
because it is not possible to
extract this information 
with a leakage scheme.

We summarise the values of luminosities
and mean energies for the three cases described
in Sec.~\ref{sec:setup} in Table \ref{Table2}.
We include the 2D values from \se,
the 2D and 3D values from M1, and the values 
from the ASL with the corresponding best parameter
set when assuming 
$\alpha_{\rm{blk}} \equiv 
\alpha_{\nu_e,\rm{blk}}= 
\alpha_{\bar{\nu}_e,\rm{blk}}$.

\begin{table*}
    \setlength{\tabcolsep}{10pt}
    \renewcommand{\arraystretch}{1.5}
	\centering
	\caption{Summary of the values of luminosities 
	and mean energies for each species from \Se
	(2nd column),
	M1 in 2D (3rd colum), M1 in 3D (4th column) and 
	the ASL (5th column), for each of the 
	three cases (1st column) examined. 
	For the ASL, we also specify the best parameter
	set resulting from the calibration (6th column) when assuming $\alpha_{\rm{blk}} \equiv \alpha_{\nu_e,\rm{blk}}= \alpha_{\bar{\nu}_e,\rm{blk}}$. 
	The reported numbers from \Se are accurate 
	in spite of the Monte Carlo 
	noise, which is estimated to be 
	a thousand times smaller.
	}
	\begin{tabular}{c|c|c|c|c|c}
		\hline
		\multicolumn{1}{c}{Case} &
		\multicolumn{1}{c}{Sedonu (2D)} &
		\multicolumn{1}{c}{M1(2D)} &
		\multicolumn{1}{c}{M1(3D)} &
		\multicolumn{1}{c}{ASL} &
		\multicolumn{1}{c}{[$\alpha_{\rm{blk}}$,$\tau_{\rm{cut}}$,$p$]}\\
		\hline
		\multirow{6}{*}{1} & $L_{\nu_e}^1= 7.50\cdot10^{51} \rm{erg\:s^{-1}}$ &
		$L_{\nu_e}^1= 7.73\cdot10^{51} \rm{erg\:s^{-1}}$ &
		$L_{\nu_e}^1= 7.46\cdot10^{51}\:\rm{erg\:s^{-1}}$
		& $L_{\nu_e}^1= 6.81\cdot10^{51}\:\rm{erg\:s^{-1}}$\\
		& $L_{\bar{\nu}_e}^1= 1.50\cdot10^{52}\:\rm{erg\:s^{-1}}$ &
		$L_{\bar{\nu}_e}^1= 1.63\cdot10^{52}\:\rm{erg\:s^{-1}}$ &
		$L_{\bar{\nu}_e}^1= 1.75\cdot10^{52}\:\rm{erg\:s^{-1}}$ 
		& $L_{\bar{\nu}_e}^1= 1.71\cdot10^{52}\:\rm{erg\:s^{-1}}$ \\
		& $L_{\nu_x}^1= 2.14\cdot10^{52}\:\rm{erg\:s^{-1}}$ &
		$L_{\nu_x}^1= 2.91\cdot10^{52}\:\rm{erg\:s^{-1}}$ &
		$L_{\nu_x}^1= 2.73\cdot10^{52}\:\rm{erg\:s^{-1}}$
		& $L_{\nu_x}^1= 2.97\cdot10^{52}\:\rm{erg\:s^{-1}}$ 
		& [0.45,10,2] \\
	    & $\langle{}E_{\nu_e}\rangle{}= 9.24\:\rm{MeV}$ & 
	    $\langle{}E_{\nu_e}\rangle{}= 9.37\:\rm{MeV}$ &
	    $\langle{}E_{\nu_e}\rangle{}= 9.38\:\rm{MeV}$
	    &  $\langle{}E_{\nu_e}\rangle{}= 8.70\:\rm{MeV}$\\
	    & $\langle{}E_{\bar{\nu}_e}\rangle{}= 13.05\:\rm{MeV}$ & 
	    $\langle{}E_{\bar{\nu}_e}\rangle{}= 13.00\:\rm{MeV}$ &
	    $\langle{}E_{\bar{\nu}_e}\rangle{}= 13.10\:\rm{MeV}$ 
	    & $\langle{}E_{\bar{\nu}_e}\rangle{}= 12.88\:\rm{MeV}$\\
	    & $\langle{}E_{\nu_x}\rangle{}= 13.87\:\rm{MeV}$ & 
	    $\langle{}E_{\nu_x}\rangle{}= 15.27\:\rm{MeV}$ &
	    $\langle{}E_{\nu_x}\rangle{}= 15.28\:\rm{MeV}$
	    &  $\langle{}E_{\nu_x}\rangle{}= 13.63\:\rm{MeV}$
	    \\ 
	    \hline
	    \multirow{6}{*}{2} & $L_{\nu_e}^1= 8.85 \cdot 10^{51}\:\rm{erg\:s^{-1}}$ &
	    $L_{\nu_e}^1= 9.03 \cdot 10^{51}\:\rm{erg\:s^{-1}}$ &
	    $L_{\nu_e}^1= 7.29\cdot10^{51}\:\rm{erg\:s^{-1}}$ &
	    $L_{\nu_e}^1= 5.65\cdot10^{51}\:\rm{erg\:s^{-1}}$\\
		& $L_{\bar{\nu}_e}^1= 8.64 \cdot 10^{51}\:\rm{erg\:s^{-1}}$ &
		$L_{\bar{\nu}_e}^1= 9.39\cdot10^{51}\:\rm{erg\:s^{-1}}$ &
		$L_{\bar{\nu}_e}^1= 1.90\cdot10^{52}\:\rm{erg\:s^{-1}}$ &
		$L_{\bar{\nu}_e}^1= 1.80\cdot10^{52}\:\rm{erg\:s^{-1}}$\\
		& $L_{\nu_x}^1= 2.00 \cdot 10^{52}\:\rm{erg\:s^{-1}}$ & 
		$L_{\nu_x}^1= 2.57\cdot10^{52}\:\rm{erg\:s^{-1}}$ &
		$L_{\nu_x}^1= 2.90\cdot10^{52}\:\rm{erg\:s^{-1}}$ &
		$L_{\nu_x}^1= 4.10\cdot10^{52}\:\rm{erg\:s^{-1}}$ 
		& [0.65,10,2] \\
	    & $\langle{}E_{\nu_e}\rangle{}= 10.65\:\rm{MeV}$ &
	    $\langle{}E_{\nu_e}\rangle{}= 10.89\:\rm{MeV}$ &
	    $\langle{}E_{\nu_e}\rangle{}= 11.11\:\rm{MeV}$ & $\langle{}E_{\nu_e}\rangle{}= 10.43\:\rm{MeV}$ \\
	    & $\langle{}E_{\bar{\nu}_e}\rangle{}= 15.74\:\rm{MeV}$
	    & $\langle{}E_{\bar{\nu}_e}\rangle{}= 15.69\:\rm{MeV}$
	    &  $\langle{}E_{\bar{\nu}_e}\rangle{}= 16.23\:\rm{MeV}$ &
	    $\langle{}E_{\bar{\nu}_e}\rangle{}= 16.61\:\rm{MeV}$ \\
	    & $\langle{}E_{\nu_x}\rangle{}= 14.01\:\rm{MeV}$ &
	    $\langle{}E_{\nu_x}\rangle{}= 14.98\:\rm{MeV}$ &
	    $\langle{}E_{\nu_x}\rangle{}= 16.05\:\rm{MeV}$ 
	    & $\langle{}E_{\nu_x}\rangle{}= 15.54\:\rm{MeV}$\\
	    \hline
	    \multirow{6}{*}{3} & $L_{\nu_e}^1= 2.04\cdot10^{52}\:\rm{erg\:s^{-1}}$ &
	    $L_{\nu_e}^1= 9.36\cdot10^{51}\:\rm{erg\:s^{-1}}$ &
	    $L_{\nu_e}^1= 6.27\cdot10^{51}\:\rm{erg\:s^{-1}}$ & $L_{\nu_e}^1= 3.01\cdot10^{51}\:\rm{erg\:s^{-1}}$\\
		& $L_{\bar{\nu}_e}^1= 2.78\cdot10^{51}\:\rm{erg\:s^{-1}}$ &
		$L_{\bar{\nu}_e}^1= 2.98\cdot10^{51}\:\rm{erg\:s^{-1}}$ &
		$L_{\bar{\nu}_e}^1= 2.24\cdot10^{52}\:\rm{erg\:s^{-1}}$ & $L_{\bar{\nu}_e}^1= 2.86\cdot10^{52}\:\rm{erg\:s^{-1}}$ \\
		& $L_{\nu_x}^1= 1.24\cdot10^{52}\:\rm{erg\:s^{-1}}$ & 
		$L_{\nu_x}^1= 1.45\cdot10^{52}\:\rm{erg\:s^{-1}}$ &
		$L_{\nu_x}^1= 2.40\cdot10^{52}\:\rm{erg\:s^{-1}}$ &
		$L_{\nu_x}^1= 3.55\cdot10^{52}\:\rm{erg\:s^{-1}}$ & [0.75,10,2]\\ 
	    & $\langle{}E_{\nu_e}\rangle{}= 6.88\:\rm{MeV}$ &
	    $\langle{}E_{\nu_e}\rangle{}= 10.60\:\rm{MeV}$ &
	    $\langle{}E_{\nu_e}\rangle{}= 9.90\:\rm{MeV}$ &
	    $\langle{}E_{\nu_e}\rangle{}= 9.61\:\rm{MeV}$\\
	    & $\langle{}E_{\bar{\nu}_e}\rangle{}= 13.17\:\rm{MeV}$
	    & $\langle{}E_{\bar{\nu}_e}\rangle{}= 14.10\:\rm{MeV}$
	    &  $\langle{}E_{\bar{\nu}_e}\rangle{}= 15.61\:\rm{MeV}$ &
	    $\langle{}E_{\bar{\nu}_e}\rangle{}= 18.85\:\rm{MeV}$\\
	    & $\langle{}E_{\nu_x}\rangle{}= 11.42\:\rm{MeV}$ & 
	    $\langle{}E_{\nu_x}\rangle{}= 12.39\:\rm{MeV}$ &
	    $\langle{}E_{\nu_x}\rangle{}= 15.56\:\rm{MeV}$ &
	    $\langle{}E_{\nu_x}\rangle{}= 16.62\:\rm{MeV}$\\ \hline
	\end{tabular}
	\label{Table2}
\end{table*}

\section{Results}
\label{sec:results}
In Sec.~\ref{sec:eachcase},
we describe the 
calibration of each ASL parameter
based on a separate analysis for 
each snapshot.
Afterwards, we combine the results in 
Sec.~\ref{sec:combination} and discuss
the performance of using the same 
blocking parameter for electron
neutrinos and anti-neutrinos. \\

\subsection{Parameter constraints}
\label{sec:eachcase}
\subsubsection{Blocking}
\begin{figure*}
\centering
\includegraphics[width = 0.95 \linewidth]{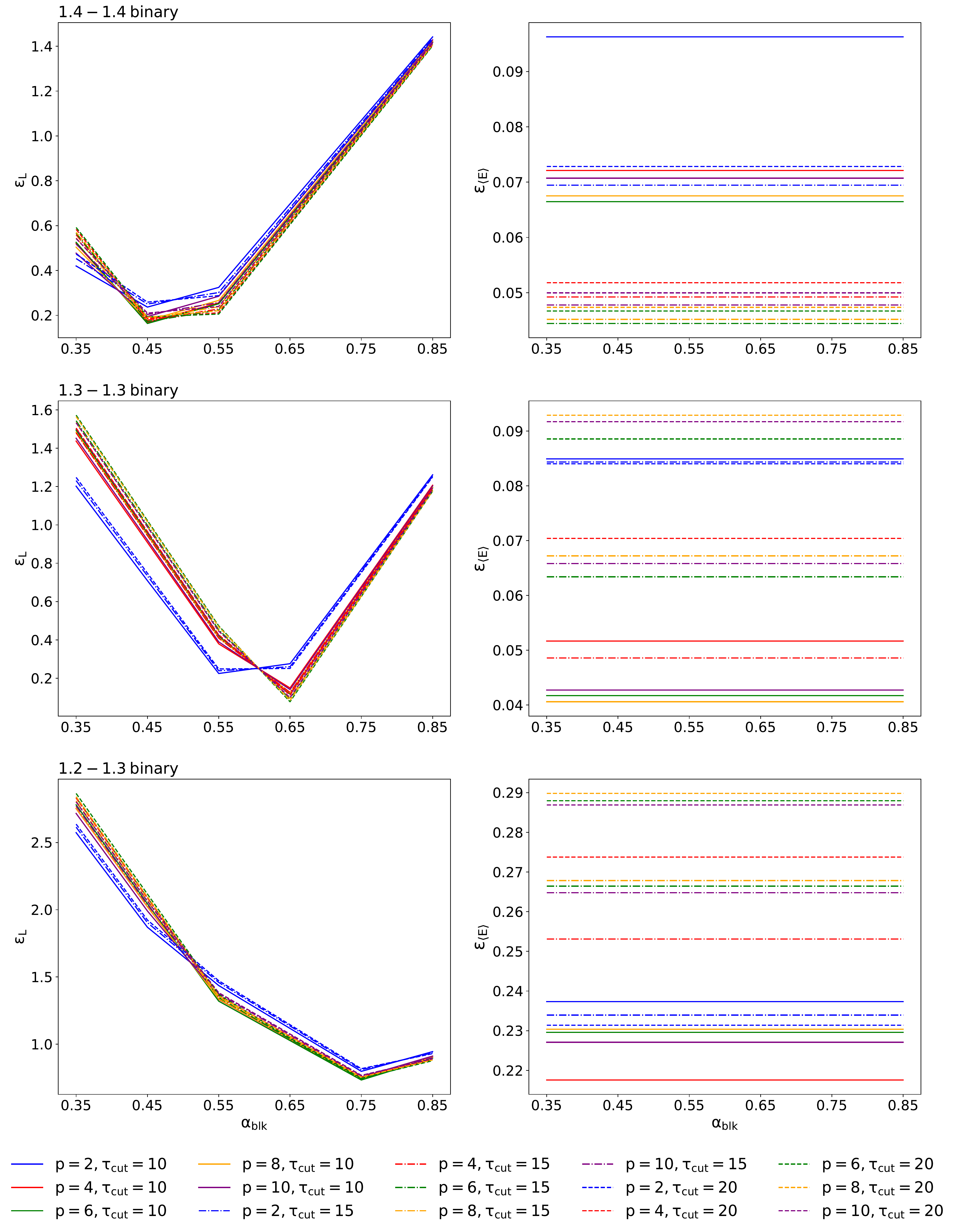}
\caption{Relative difference on the 
luminosity $\epsilon_L$ (left panels)
and the mean neutrino energy 
$\epsilon_{\langle E \rangle}$ (right panels)
as a function of $\alpha_{\rm{blk}}$ and
for different values of 
$p$ and $\tau_{\rm{cut}}$, calculated 
by comparing the results from the 
ASL with the reference 
solution. From top to bottom, we show
cases from 1) to 3). We use 
\Se as reference solution for 
case 1), and M1 for cases 2) and 3).
Lines of the same color are for 
fixed $p$, while lines of the same 
type are for fixed $\tau_{\rm{cut}}$.
We neglect the cases with $\tau_{\rm{cut}}=5$
to reduce the amount of data to show. 
The parameter $\alpha_{\rm{blk}}$ 
has a large impact on 
$\epsilon_L$, but not on 
$\epsilon_{\langle E \rangle}$.
The thermalization parameter 
$\tau_{\rm{cut}}$
has some effect on 
$\epsilon_{\langle E \rangle}$,
while $p$ mildly impacts both 
$\epsilon_L$ and 
$\epsilon_{\langle E \rangle}$.}
\label{fig:epsLE}
\end{figure*}

From Table \ref{Table2} we can see
that by comparing the 2D and 3D 
luminosities of electron
neutrinos and anti-neutrinos from M1 
the impact of the 2D averaging is
only of the order of 
$\lesssim 7\%$ for both 
species in case 1).
This suggests the usage of 
\Se for the comparison 
with the ASL scheme. 
On the other hand,
for cases 2) and 3) the
2D averaging implies a reduction in the
electron anti-neutrino luminosity
by about a factor of 2 
and 10 respectively. 
We examine the effect of 
the 2D averaging on
the emission maps for each species, 
and we indeed
notice that the 2D assumption
affects the neutrino species 
to a different
extent, depending on the respective 
location of 
the bulk of the emission.
We therefore decide to 
take the 3D M1 values
of luminosities and mean energies 
as reference
for calibrating $\alpha_{\rm{blk}}$ and 
$\tau_{\rm{cut}}$ in cases 2) and 3).
We also notice that 
the heavy-lepton
neutrino luminosity from M1 in 3D
and from the ASL 
is systematically larger 
by a factor of a few
than the one from \Se in all snapshots.
This is different from
the results of 
\cite{Foucart2020}.
However, the fact that only
heavy-lepton neutrinos show
large deviations with respect to a 
Monte Carlo approach 
is an indication that
the cause might be associated to
some ingredient in the
particular transport approach
adopted, and for which heavy-leptons
are more affected than the other species.
Regarding the ASL, a likely
explanation is the treatment of the 
emission rate, which is calculated
as smooth interpolation between
production and diffusion rates
as shown in Eq.~\ref{eq:ser}. 
The diffusion rate depends on the 
diffusion timescale, which is 
estimated via a random-walk argument.
As pointed out in \cite{Ardevol19}, 
this derivation leads
to a steeper decrease of the diffusion
timescale with radius. 
Combined with the fact that 
most of the neutrinos escape
around the rather small region 
of the neutrino surface, 
a lower diffusion timescale in this
region can boost the emission up to 
more than a factor of 2, 
depending on the species. 
For our binary neutron star configurations,
this is particularly true for heavy-lepton
neutrinos because their 
sources of production are just
pair processes and bremsstrahlung, 
which are both extremely 
temperature-dependent. Combined with 
the fact that 
heavy-lepton neutrinos 
decouple at inner and 
still rather hot regions
with respect to the other 
two species, it is likely 
that their emission is 
more affected by the 
treatment of the diffusion.\\
Fig.~\ref{fig:epsLE} shows 
line plots of $\epsilon_{L}$ 
and $\epsilon_{\langle E \rangle}$
as a function of $\alpha_{\rm{blk}}$,
for different values of $p$ and $\tau_{\rm{cut}}$. 
Lines of the same color are for 
fixed $p$, while lines of the same 
type are for fixed $\tau_{\rm{cut}}$.
We neglect the cases with $\tau_{\rm{cut}}=5$
and $p=12$
to reduce the amount of data to show,
but the results do not change.
From the top to the bottom row 
we show the cases from 1) to 3)
respectively.
Generally, we see that 
neither $\epsilon_{L}$
nor $\epsilon_{\langle E \rangle}$ 
is largely affected
by varying $p$ for a 
given $\alpha_{\rm{blk}}$ 
and $\tau_{\rm{cut}}$. 
The mild dependence can be 
explained by the fact that $p$ defines 
the distribution of the heating rather 
than its intensity.
However, because the neutrino 
absorption is typically
more pronounced for 
electron neutrinos 
than for anti-neutrinos,
we should expect a somewhat larger
dependence of
$\epsilon_{L}$ 
on $p$ for the former species.
Nevertheless, since we are evaluating the 
species-summed $\epsilon_{L}$, 
even in case there is such dependence
from the electron neutrinos, this 
must be associated to a rather small 
$\epsilon_{L_{\nu_e}}$ 
(see later Fig.~\ref{fig:epsLiEi}), 
and it is thus not appreciable. 
On the other hand, the 
blocking parameter $\alpha_{\rm{blk}}$
has a major impact on $\epsilon_{L}$. This comes
directly from the $r_{\nu} \rightarrow (1-\alpha_{\rm{blk}})r_{\nu}$ correction 
to the emission rate $r_{\nu}$ when 
accounting for blocking, see
Eq.~\ref{eq:ser}. 
This is different from the 
impact on $\epsilon_{E}$, 
which is basically negligible.
The reason is the fact that 
$\langle E_{i} \rangle$
is computed from the ratio between 
the luminosity $L_{i}$ [erg/s] and 
the total number of emitted neutrinos of 
species $i$ per unit time, both affected 
by blocking for the case of electron
neutrinos and anti-neutrinos.
At last, a noticeable dependence to
 $\tau_{\rm{cut}}$ can 
be seen from $\epsilon_{\langle E \rangle}$, 
as a consequence of the 
fact that $\tau_{\rm{cut}}$ impacts 
the neutrino spectrum
at the decoupling region, and 
therefore the mean energies of the 
species. We find that 
$\epsilon_{\langle E \rangle}$
has a maximum value of 
$\sim 18\%$ and $\sim 12\%$ 
at $\tau_{\rm{cut}}=5$
for cases 1) and 2) respectively.
Case 3) shows larger values, but 
limited to $< 30\%$.
On the other hand, 
$\epsilon_L$ varies from
$\sim 10\%$ to more than $100\%$
for cases 1) and 2), while 
case 3) shows values 
$\epsilon_{L} \gtrsim 0.75$.
The large $\epsilon_{L}$
for any parameter
combination for the latter case is 
a consequence of the 
$\alpha_{\rm{blk}}=
\alpha_{\nu_e,\rm{blk}}= 
\alpha_{\bar{\nu}_e,\rm{blk}}$
assumption, which makes it
cumbersome to always 
well catch the
luminosities of both 
electron neutrinos and
anti-neutrinos (see 
also Fig.~\ref{fig:epsLiEi}
and next paragraph).
Since $\epsilon_{L}$
is more sensitive
than $\epsilon_{\langle E \rangle}$
to a change in 
$\alpha_{\rm{blk}}$, 
we just
look at $\epsilon_{L}$ 
for a first parameter pre-selection.
In particular, we find 
a minimum $\epsilon_{L}$ around
$\alpha_{\rm{blk}}=0.45$ for case 1),
$\alpha_{\rm{blk}}=0.65$ for case 2),
and $\alpha_{\rm{blk}}=0.75$ for case 3).

\subsubsection{Thermalization}

\begin{figure*}
\centering
\includegraphics[width = 0.95 \linewidth]{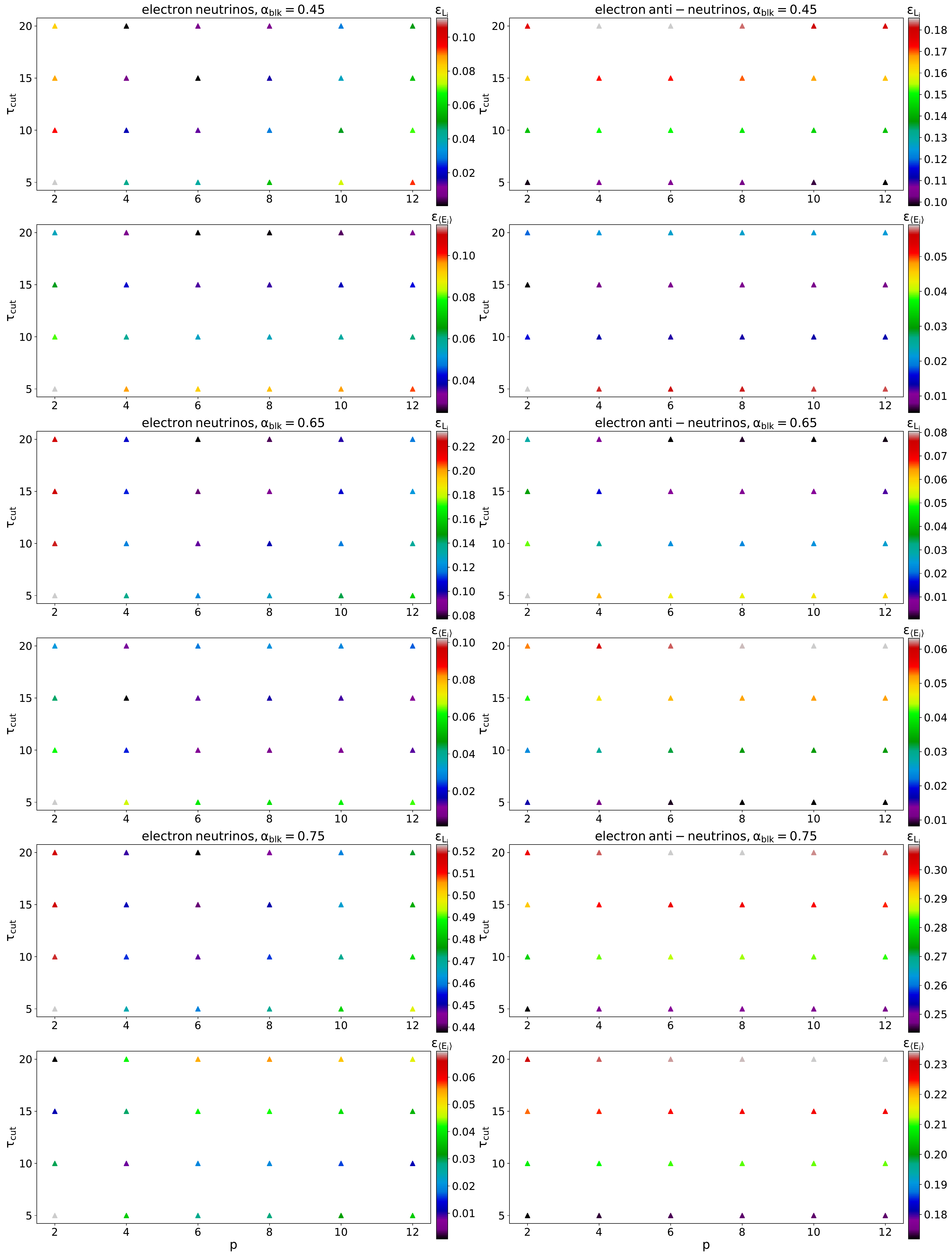}
\caption{Relative difference on the luminosity 
$\epsilon_{L_i}$
and the mean energy 
$\epsilon_{\langle E_i \rangle}$
for $i=\nu_e$ (left panels) 
and $i=\bar{\nu}_e$ (right panels), and 
for case 1) (first and second row), 
case 2) (third and fourth row), and 
case 3) (fifth and sixth row), calculated 
by comparing the results from the ASL 
with the reference solution.
We use \Se as reference solution for 
case 1), and M1 for cases 2) and 3).
While $\epsilon_{\langle E_{i} \rangle}$
is always limited to $\lesssim 23\%$ 
at the most for electron anti-neutrinos 
in case 3), the assumption 
of a single blocking parameter
for both electron neutrinos
and anti-neutrinos leads to 
inaccurate recoveries of the neutrino
luminosities in some cases, 
with $\epsilon_{L_{\nu_e}}$ up to $52\%$.}
\label{fig:epsLiEi}
\end{figure*}

We set $\alpha_{\rm{blk}}$ 
to the above values, and 
look at both $\epsilon_{L_i}$ and 
$\epsilon_{\langle E_i \rangle}$ for 
each species $i \in [\nu_e,\bar{\nu}_e]$
in Fig.~\ref{fig:epsLiEi}.
The first two rows correspond to 
the results for case 1), the third and 
fourth row to the results for case 2),
and the last two rows for case 3).
We show both results for electron neutrinos
(left column) and for electron anti-neutrinos
(right column). As anticipated before, 
$\epsilon_{L_{\nu_e}}$
shows the
largest dependence on $p$ for a
given $\tau_{\rm{cut}}$.
We find that for electron neutrinos 
$\epsilon_{L_{\nu_e}} \lesssim 12\%$,
$\epsilon_{L_{\nu_e}} \lesssim 23\%$ and
$\epsilon_{L_{\nu_e}} \lesssim 52\%$ for 
cases 1), 2) and 3) respectively.
If we look at electron anti-neutrinos we have
$\epsilon_{L_{\bar{\nu}_e}}$ up to $\sim 20\%$,
$\epsilon_{L_{\bar{\nu}_e}} \lesssim 8\%$
and $\epsilon_{L_{\bar{\nu}_e}} \lesssim 31\%$
respectively.
On the other hand, 
$\epsilon_{\langle E_i \rangle}
\lesssim 12\%$ for each species $i$
in cases 1) and 2), while 
case 3) shows 
$\epsilon_{\langle E_{\bar{\nu}_e} \rangle}
\lesssim 23\%$.
While the overall 
limited values 
in both $\epsilon_{L_i}$ and 
$\epsilon_{\langle E_i \rangle}$ 
suggest that 
any value of $\tau_{\rm{cut}}
\in [5,10,15,20]$ seems good enough
to describe the thermalization
for both cases 1) and 2), 
in case 3) only either 
$\tau_{\rm{cut}}= 5$
or $\tau_{\rm{cut}}= 10$
is able to keep 
$\epsilon_{L_{\bar{\nu}_e}} < 30\%$,
in spite of the large
$\epsilon_{L_{\nu_e}} \gtrsim 52\%$.
We therefore always take $\tau_{\rm{cut}}= 10$ 
as reference for the ASL luminosities 
and mean energies
shown in Table \ref{Table2}.\\
We notice that
the electron neutrino and anti-neutrino 
mean energies are systematically higher 
for cases 2) and 3)
in both the M1 and the ASL
by up to $\sim 1-3$ MeV
with respect to case 1).
We provide an explanation 
by calculating the average temperatures 
at which both neutrino species are emitted
by means of equation 9 of
\cite{Rosswog2003}.
Table \ref{Table3} shows 
the results we find.
The average temperatures 
are higher by $\gtrsim 1$ MeV 
with respect to case 1), implying that
most of the emission for both electron 
neutrinos and anti-neutrinos comes from
hotter regions. 
Since neutrinos thermalize with matter 
before free-streaming, the mean neutrino
energy is higher if temperatures are higher.
The presence
of hotter regions is also confirmed 
by the fact that the maximum temperature
seen for case 1) is $\sim 25$ MeV, while
for case 2) reaches $\sim 40$ MeV and 
for case 3) $\sim 32$ MeV.
However, the higher electron 
anti-neutrino mean energy in case 3) is 
also a consequence of the rather large 
$\epsilon_{\bar{\nu}_e}$
(Fig.~\ref{fig:epsLiEi}, right panel 
on the sixth row).
\subsubsection{Heating}
\begin{figure*}
\centering
\includegraphics[width = 0.82 \linewidth]{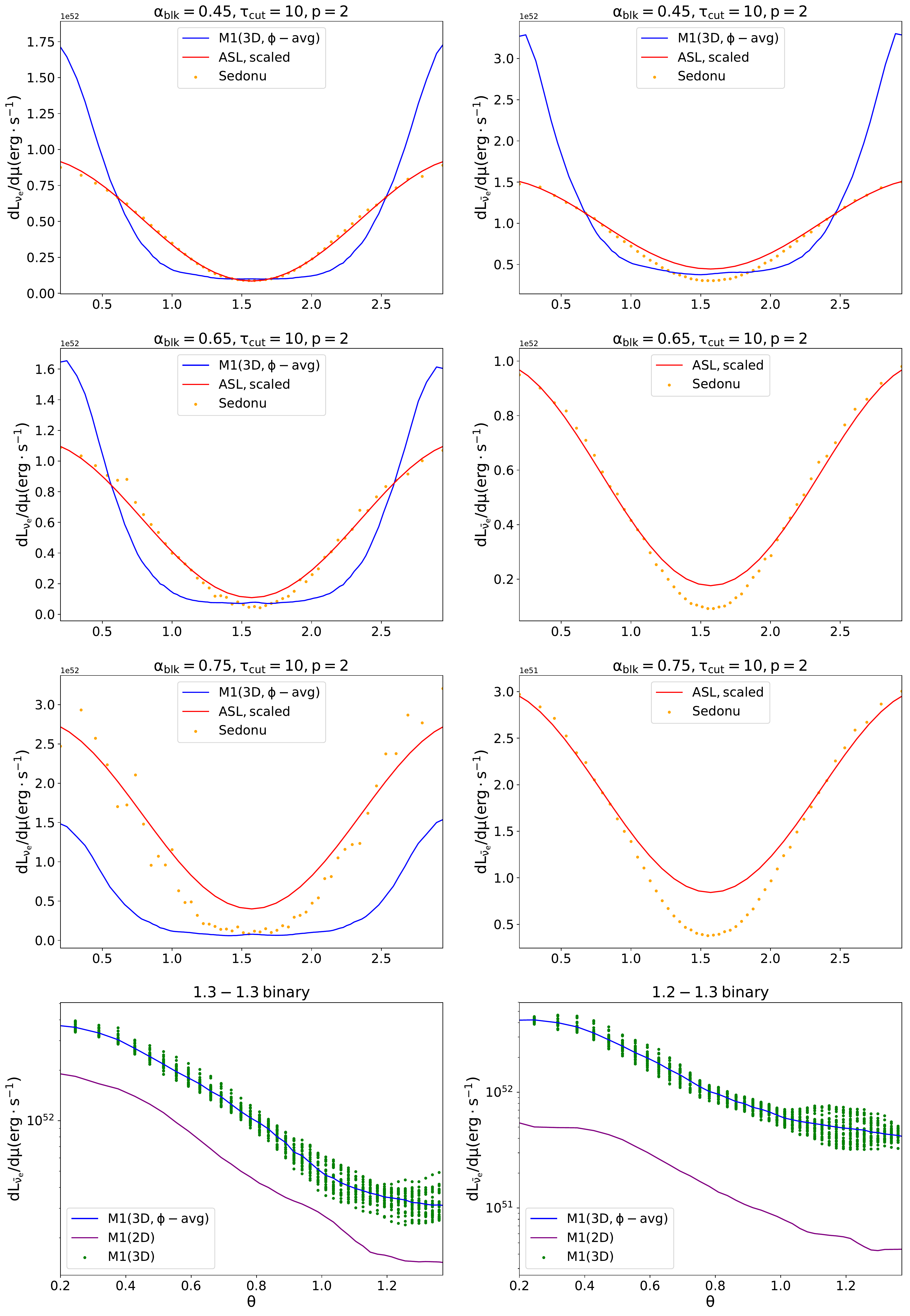}
\caption{$dL_{i}/d \mu$ as a function of $\theta$.
From the first to the third row
we show the results for cases 
from 1) to 3) respectively,
for $i=\nu_e$ (left)
and $i=\bar{\nu}_e$ (right).
Dots represent
the exact solution from \se.
The blue curve describes the trend of
$dL_{i}/d \mu$
obtained with M1 when performing
an azimuthal average over different
$\phi$ at each $\theta$, 
while the red curve describes the 
one from the ASL, scaled
by some constant, and obtained via
Eq.~\ref{eq:flux2} with
the best parameter set of each case. 
We only show the trends from 
\Se and ASL 
for anti-neutrinos
in cases 2) and 3).
The trend from \Se is overall well described by a 
$\rm{cos}^{2}(\theta)$ for both species
and in all cases. 
M1 tends to overestimate the flux 
close to the polar axis
for both electron neutrinos and anti-neutrinos,
as a consequence of the
approximations introduced by 
the analytical closure.
The last row shows 
$dL_{\bar{\nu}_e}/d \mu$
as a function of $\theta$ for
case 2) (left), and 
case 3) (right).
The 2D assumption does not
affect the trend of $dL_{\bar{\nu}_e}/d \mu$.
Moreover, for a given $\theta$
we see a limited 
variation of $dL_{\bar{\nu}_e}/d \mu$
with $\phi$ at polar angles 
relevant for neutrino-driven winds
($\theta \lesssim \pi/3$).}
\label{fig:dLdmu}
\end{figure*}

To constrain the heating 
parameter $p$, we show
in the first three 
rows of Fig.~\ref{fig:dLdmu}
$dL_{i}/d\mu$ as a function
of $\theta$ for all cases examined, 
and for $i=\nu_e$ (left panels),
$i=\bar{\nu}_e$ (right panels).
The curve from M1 is
obtained by taking an average of $dL_{i}/d\mu$
over different azimuthal 
angles $\phi$ of the domain 
for each polar angle $\theta$, while
the ASL curve is obtained by means 
of Eq.~\ref{eq:flux2},
with the best blocking parameter,
a reference value of $\tau_{\rm{cut}}= 10$
\footnote{We just use this 
value of $\tau_{\rm{cut}}$ to illustrate
the calibration of $p$. 
Other choices of $\tau_{\rm{cut}}$
would have been equivalent.}, and 
the best $p$ from the 
direct comparison between 
the polar angle profiles of the luminosities. 
Unlike case 1), we only show
the 2D result from \Se and 
the ASL curve of $dL_{\bar{\nu}_e}/d \mu$
for cases 2) and 3) because of the
impact of the 2D assumption on 
the electron anti-neutrino emission.
We see that $p=2$ 
nicely fits trends from \Se for both 
electron neutrinos and anti-neutrinos
and in all cases.
We can also see that M1 
overestimates the flux 
close to the polar axis
for both electron neutrinos and
anti-neutrinos in cases 1) and 2),
with a maximum disagreement 
with respect to the result from 
\Se by a factor of $\lesssim 2$ at the pole
in case 1).
This is in accordance with the
previous findings of \cite{Foucart2018,Foucart2020},
and it is the result of the 
approximations introduced by
the analytical closure.
It is also important to notice
that our new estimate of the 
parameter $p=2$ is lower 
than the one we found in
\cite{Gizzi2019}, where we estimated
$p=8$. The reason is the fact 
that there we 
chose M1 as source for comparison.
As just said, 
the approximations introduced by
the analytical closure inevitably
leads to a steeper decrease of the
neutrino fluxes with the polar angle,
consequently suggesting a larger
value of $p$ than the one found here.
We justify our decision of using 
the 2D data from \Se of 
electron anti-neutrinos 
in cases 2) and 3) for the
calibration of $p=2$ by noticing that
the 2D assumption does not impact 
the trend of $dL_{\bar{\nu}_e}/d\mu$ in M1
(see purple curve in the last row
of Fig.~\ref{fig:dLdmu}).
The last row shows
also the quantity $dL_{\bar{\nu}_e}/d\mu$
as function of the polar angle 
$\theta$ for different azimuthal
angles $\phi$, obtained with M1 in 3D
(green dots). 
We can clearly see that
the variation with $\phi$ is limited, 
particularly at angles 
$\theta \lesssim \pi/3 \approx 1.05$ 
where the bulk of neutrino-driven winds is located.
This justifies the assumption of axially symmetric
fluxes entering Eq.~\ref{eq:new_nudensity}.

\begin{table*}
\setlength{\tabcolsep}{10pt}
\renewcommand{\arraystretch}{1.5}
\centering
\caption{Average densities (2nd and 3rd columns), 
temperatures (4th and 5th columns) and  
electron fractions (6th and 7th columns)
at which electron neutrinos
and anti-neutrinos are emitted, 
and difference between average electron 
anti-neutrino and electron neutrino
degeneracy parameters (8th column),
for each of the three cases examined (1st column).
The computation of the averages 
follows equation 9 of \protect\cite{Rosswog2003}.
For the computation of the neutrino chemical 
potentials entering the degeneracy parameters, 
we assume weak equilibrium.}
    \begin{tabular}{c|c|c|c|c|c|c|c}
      \hline
      Case & $\langle \rho \rangle_{\nu_e}$[$\rm{10^{13}\:g\cdot cm^{-3}}$] & $\langle \rho \rangle_{\bar{\nu}_e}$[$\rm{10^{13}\:g\cdot cm^{-3}}$] & $\langle T \rangle_{\nu_e}$[MeV] & $\langle T \rangle_{\bar{\nu}_e}$[MeV] & $\langle Y_e \rangle_{\nu_e}$ & $\langle Y_e \rangle_{\bar{\nu}_e}$ & $\langle \eta \rangle_{\bar{\nu}_e} - \langle \eta \rangle_{\nu_e}$\\ \hline 
      1 & 0.86 & 1.77 & 4.37 & 5.46 & 0.075 & 0.070 & 1.38\\ \hline
      2 & 1.18 & 1.03 & 5.51 & 6.55 & 0.104 & 0.091 & 0.95\\ \hline
      3 & 0.80 & 0.52 & 5.16 & 6.41 & 0.085 & 0.073 & 1.93\\ \hline
    \end{tabular}
    \label{Table3}
\end{table*}
\subsection{Combination of parameter constraints}
\label{sec:combination}

\begin{figure*}
\centering
\includegraphics[width = 0.98 \linewidth]{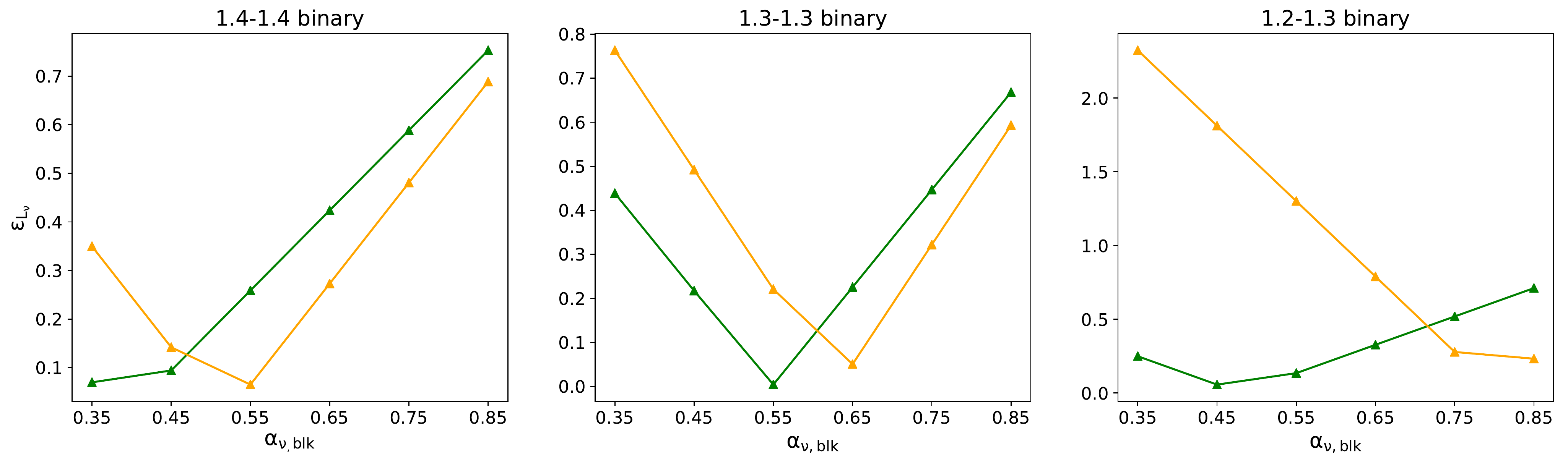}
\caption{$\epsilon_{L_{\nu}}$ as a 
function of 
$\alpha_{\nu,\rm{blk}} \in 
[0.35,0.45,0.55,0.65,0.75,0.85]$, 
for the three binary setups and 
for both electron neutrinos (green curve)
and anti-neutrinos (orange curve).
Plots are obtained with $\tau_{\rm{cut}}=10$
for all binaries, and with $p=2$ and 
$p=4$ for equal and unequal mass binaries 
respectively.
For each binary, the 
value of the blocking 
corresponding to the minimum 
$\epsilon_{L_{i}}$ is 
lower for $i=\nu_e$ than for
$i=\bar{\nu}_e$. While 
for $\alpha_{\nu_e,\rm{blk}}= 0.45$
$\epsilon_{L_{\nu_e}} \lesssim 20\%$
in all binaries, for the
anti-neutrinos the larger variability
of $\epsilon_{L_{\bar{\nu}_e}}$ with 
$\alpha_{\bar{\nu}_e,\rm{blk}}$ when
exploring different cases leads to 
a different fiducial value for 
equal and unequal mass binaries.
In particular, 
$\alpha_{\bar{\nu}_e,\rm{blk}}=0.55$
for the former case, and 
$\alpha_{\bar{\nu}_e,\rm{blk}}=0.75$
for the latter case. 
In this way, 
$\epsilon_{L_{\bar{\nu}_e}} \lesssim 20\%$
and $\epsilon_{L_{\bar{\nu}_e}} < 30\%$
respectively.
}
\label{fig:epsLnui}
\end{figure*}

\begin{figure*}
\begin{minipage}{1 \linewidth}
\centering
\includegraphics[width = 1 \linewidth]{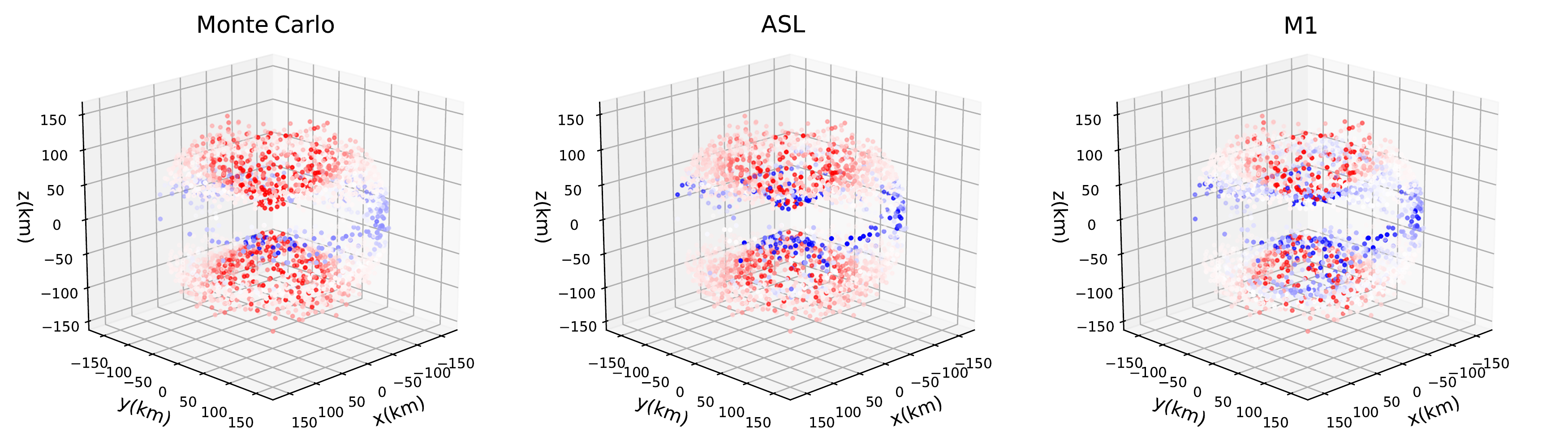} 
\end{minipage} \\
\begin{minipage}{1 \linewidth}
\centering
\includegraphics[width = 1 \linewidth]{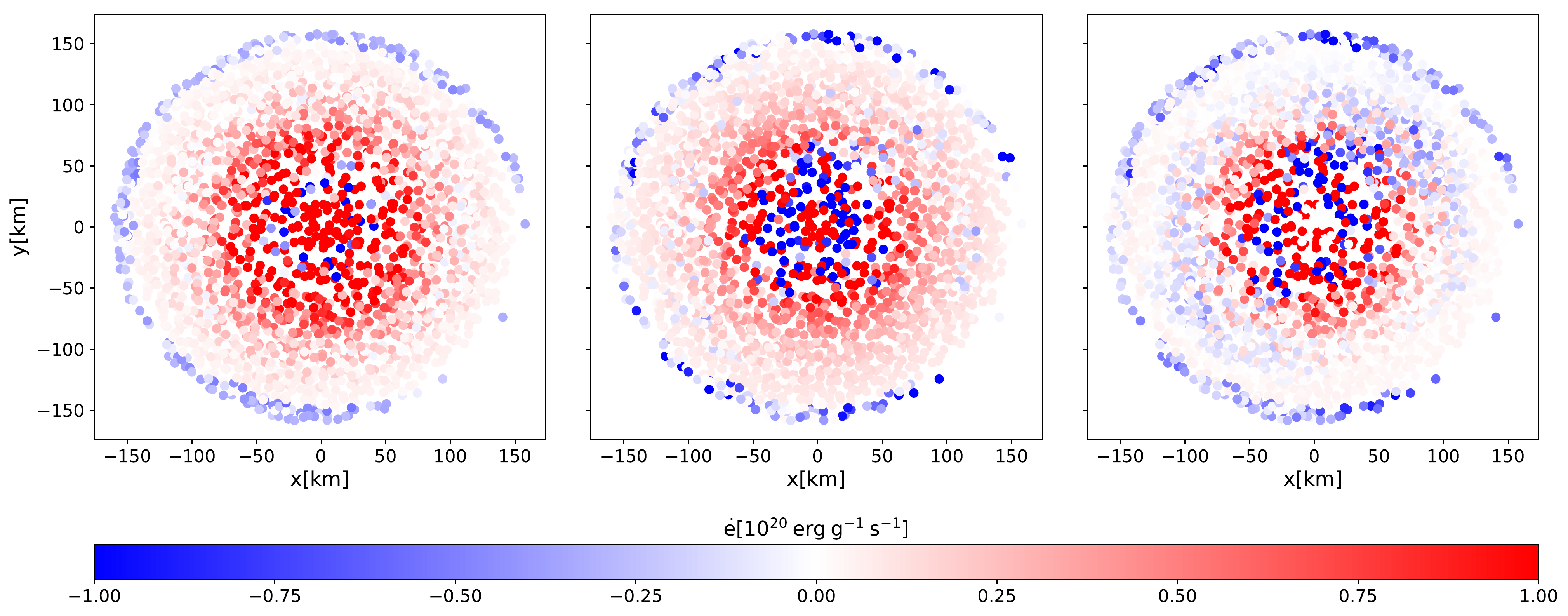}
\end{minipage}
\caption{3D maps (top row) and 
projection of the 3D maps 
on the x-y plane (bottom row)
of the rate of change 
of the internal energy 
$\dot{e}$ in the winds (units of $10^{20}\:\rm{erg}\:\rm{g}^{-1}\:\rm{s}^{-1}$) 
for \Se (left column),
the ASL (middle column) and M1 (right column), case 1).
The ASL performance is similar to M1, 
and they both generally reproduce the distribution 
of $\dot{e}$ from \se.
The main difference with respect to \Se
is in somewhat lower $\dot{e} > 0$ and 
stronger $\dot{e} < 0$ locally.
The domain size is $x \times y \times z = 
150\:\rm{km} \times 150\:\rm{km} \times 150\:\rm{km}$.
The map for the ASL is obtained with
the parameter set 
$\alpha_{\nu_e,\rm{blk}}= 0.45,
\alpha_{\bar{\nu}_e,\rm{blk}}= 0.55, 
\tau_{\rm{cut}}= 10, p=2$.}
\label{fig:edot1414}
\end{figure*}

\begin{figure*}
\begin{minipage}{1 \linewidth}
\centering
\includegraphics[width = 1 \linewidth]{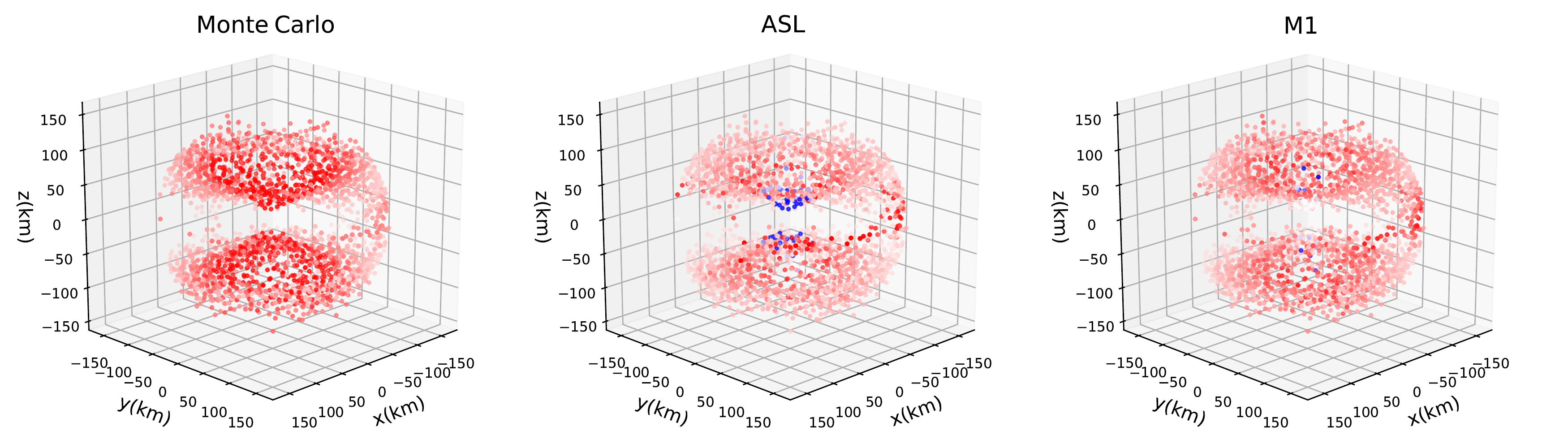}
\end{minipage}\\
\begin{minipage}{1 \linewidth}
\centering
\includegraphics[width = 1 \linewidth]{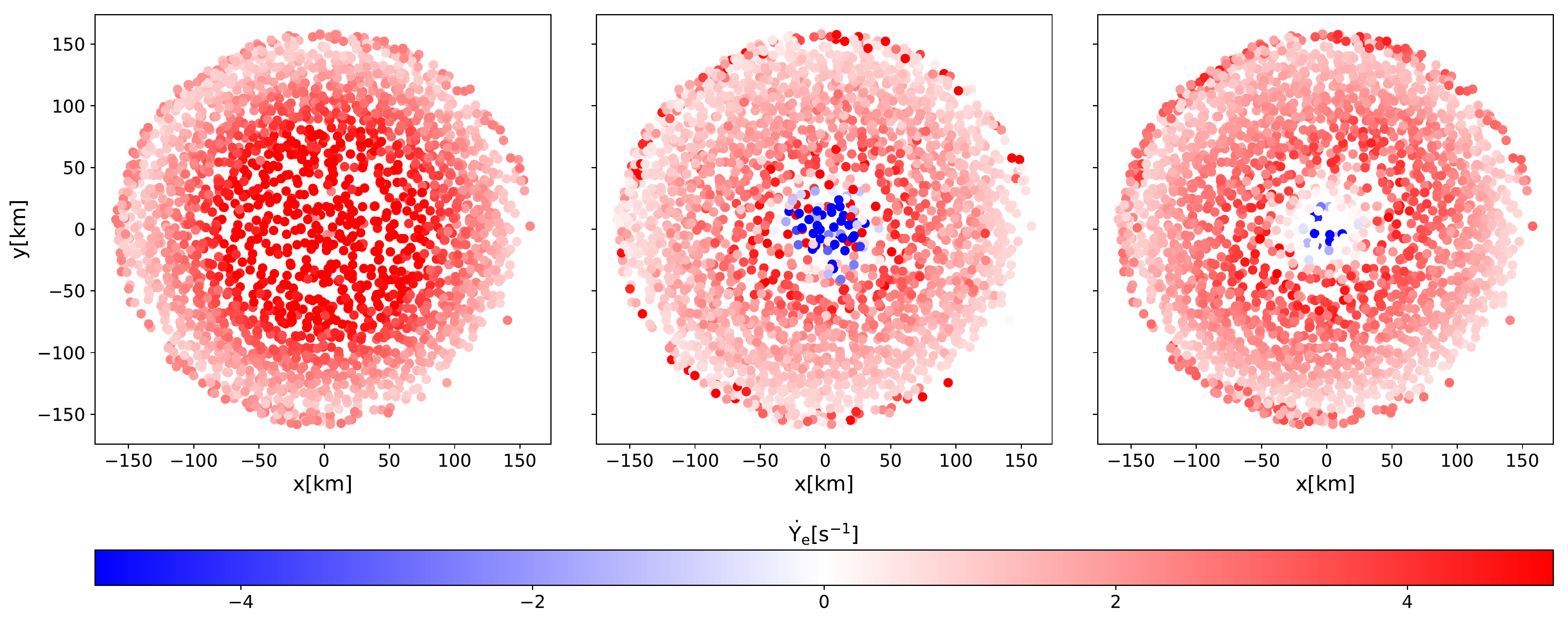}
\end{minipage}
\caption{3D maps (top row) and 
projection of the 3D maps on the x-y plane
(bottom row) of the rate of change
of the electron fraction $\dot{Y}_{\rm{e}}$ 
in the winds for \Se (left column),
the ASL (middle column) and M1 (right column).
The ASL performance is very similar to M1, 
and they both generally reproduce the distribution 
of $\dot{Y}_{\rm{e}}$ from \se. 
The latter provides always $\dot{Y}_{\rm{e}} > 0$,
while both the ASL and M1 show a few particles with
$\dot{Y}_{\rm{e}} < 0$. Moreover, 
$\dot{Y}_{\rm{e}} > 0$
from \Se appears stronger
than in the ASL and M1.
The domain size is $x \times y \times z = 
150\:\rm{km} \times 150\:\rm{km} \times 150\:\rm{km}$.
The map for the ASL is obtained with
the parameter set 
$\alpha_{\nu_e,\rm{blk}}= 0.45,
\alpha_{\bar{\nu}_e,\rm{blk}}= 0.55, 
\tau_{\rm{cut}}= 10, p=2$.}
\label{fig:yedot1414}
\end{figure*}

Among the three calibrated parameters, 
both $\tau_{\rm{cut}}$ and $p$ do not 
show variations 
by changing the binary 
configuration. In particular,
$\tau_{\rm{cut}}$ is
always  around a value of 10,
and $p=2$. On the other hand,
$\alpha_{\rm{blk}}$ is more 
sensitive than the other parameters
when moving from one binary configuration to 
another.
Similar to the results of \cite{Perego2016},
the blocking parameter may vary 
in a range [0.45,0.75], depending
on the configuration of the binary
and its time after merger.
Under the assumption 
$\alpha_{\rm{blk}} \equiv
\alpha_{\nu_e,\rm{blk}}=\alpha_{\bar{\nu_e},\rm{blk}}$,
we find larger values of $\alpha_{\rm{blk}}$
for cases 2) and 3) with respect to case 1).
Specifically,
$\alpha_{\rm{blk}}$ is 
$\sim 44\%$ and $\sim 66\%$
larger respectively.
However, it is important to 
consider that the values of 
luminosities and mean energies 
from M1 that we 
have taken as reference
in cases 2) and 3)
might be off by $\sim 20\%$
with respect to an exact solution
to the transport
\citep{Foucart2020}, implying that
our calibrated $\alpha_{\rm{blk}}$
might be slightly affected too.\\
We find that assuming $\alpha_{\nu_e,\rm{blk}}=
\alpha_{\bar{\nu_e},\rm{blk}}$
is not a good choice at early 
times after merger.
In particular, we notice that 
higher values of $\alpha_{\rm{blk}}$
do not well capture the total 
luminosity of electron neutrinos
(see left panels in the 
third and fifth row of
Fig.~\ref{fig:epsLiEi}), pointing to the need
for $\alpha_{\nu_e,\rm{blk}} < \alpha_{\bar{\nu}_e,\rm{blk}}$. This conclusion is consistent 
with the fact that electron anti-neutrinos
are the most emitted species
in merger environments, and therefore
likely more affected by blocking effects.
Moreover, we find the 
electron anti-neutrino
gas to have on average 
fewer energy states 
available to be populated 
by new emitted neutrinos
of the same species. 
Indeed, by calculating the difference
between the average degeneracy parameter
\footnote{The degeneracy parameter is locally defined as $\eta_i=\mu_i/T$, where $\mu_i$ is the chemical potential
of neutrino of species $i$ and $T$ is the temperature.}
of electron neutrinos and anti-neutrinos,
we find positive values for all 
the cases examined (see 8th column 
of Table~\ref{Table3}).
In light of this, 
we optimize the choice of 
blocking parameter 
by exploring in Fig~\ref{fig:epsLnui} 
$\epsilon_{L_{i}}$ 
for $i=[\nu_e,\bar{\nu}_e]$ and
for the three cases examined 
in Sec.~\ref{sec:calibration}
by assuming $\tau_{\rm{cut}}=10$, 
$p=2$
and $\alpha_{\nu_e,\rm{blk}} \neq \alpha_{\bar{\nu_e},\rm{blk}}$.
We confirm that for each binary the 
value of the blocking 
corresponding to the minimum 
$\epsilon_{L_{i}}$ is 
lower for $i=\nu_e$ than for
$i=\bar{\nu}_e$.
For electron neutrinos, the 
largest blocking parameter 
at which $\epsilon_{L_{\nu_e}}$
is minimum is found in
case 2), with 
$\alpha_{\nu_e,\rm{blk}}= 0.55$. 
By looking at Table \ref{Table3},
this is on one side due to
the larger average emission temperature
(${\langle T \rangle}_{\nu_e} = 5.51\:\rm{MeV}$), 
for which
the emission rate is enhanced considerably
due to the $\sim T^5$ dependence
for charged-current interactions 
\citep{Rosswog2003}, and on the
other side to the less 
neutron-rich environment 
(${\langle Y_e \rangle}_{\nu_e}= 0.104$)
that favours electron captures on protons.
Both factors contribute in 
providing more electron neutrinos, 
and therefore
enhancing Pauli blocking effects.
For electron anti-neutrinos
the largest blocking parameter
at which $\epsilon_{L_{\bar{\nu}_e}}$
is minimum is found for case 3), with 
$\alpha_{\bar{\nu}_e,\rm{blk}}= 0.75-0.85$.
Again, this is due to both a
rather hot
(${\langle T \rangle}_{\bar{\nu}_e}= 6.41\:\rm{MeV}$)
and neutron-rich material 
(${\langle Y_e \rangle}_{\nu_e}= 0.073$)
that favours positron captures on 
neutrons.
A value of $\alpha_{\nu_e,\rm{blk}}= 0.45$
provides $\epsilon_{L_{\nu_e}} \lesssim 20\%$
for all binaries, and we therefore set it  
as fiducial for this species.
Regarding the anti-neutrinos, 
the variability of 
$\epsilon_{L_{\bar{\nu}_e}}$
with $\alpha_{\bar{\nu}_e,\rm{blk}}$
is larger and makes the choice
of the best blocking parameter 
more cumbersome.
In particular, equal mass 
binaries prefer 
$\alpha_{\bar{\nu}_e,\rm{blk}}= 0.55$
and $\alpha_{\bar{\nu}_e,\rm{blk}}= 0.65$,
for cases 1) and 2) respectively, while
case 3) prefers $\alpha_{\bar{\nu}_e,\rm{blk}} > 0.65$.
We therefore suggest a fiducial value of
$\alpha_{\bar{\nu}_e,\rm{blk}}= 0.55$ for
equal mass binaries, such that 
$\epsilon_{L_{\bar{\nu}_e}} \lesssim 20\%$,
and $\alpha_{\bar{\nu}_e,\rm{blk}}= 0.75$
for unequal mass binaries, such that 
$\epsilon_{L_{\bar{\nu}_e}} < 30\%$.
However, the unequal mass case might 
need to be explored in other test cases
for a more robust gauging of 
$\alpha_{\bar{\nu}_e,\rm{blk}}$.
Beside the thermodynamical,
compositional and degeneracy
properties of the matter described in 
Table \ref{Table3} and determining the
extent of Pauli blocking, the increasing 
value of the blocking parameter 
for both electron neutrinos 
and anti-neutrino species
when moving from case 1) to case 3) 
is also consistent with the fact 
that generally less massive and/or
unequal mass binaries produce 
larger disks 
\citep{Rosswog2000,Vincent2019,Bernuzzi2020b}. 
As stated in Sec.~\ref{sec:ASL},
the blocking parameter takes also into
account the reduction of neutrino 
emission due to inward neutrino 
fluxes in the semi-transparent regime.
The presence of larger disks 
in less massive and/or
unequal mass binaries
leads to larger neutrino surfaces, 
and consequently to an overall
larger effect of inward neutrino fluxes,
therefore contributing to 
an increase in the size of blocking.\\
We conclude by showing 
in Figs.~\ref{fig:edot1414}-\ref{fig:yedot1414}
the distribution of the  
rate of change of
the specific matter internal energy 
$\dot{e}$ (units of $10^{20}\:\rm{erg}\:\rm{g}^{-1}\:\rm{s}^{-1}$)
and of the electron
fraction $\dot{Y}_{\rm{e}}$
in the winds for case 1) and for 
the different transport approaches, 
assuming $\alpha_{\nu_e,\rm{blk}}= 0.45,
\alpha_{\bar{\nu}_e,\rm{blk}}= 0.55, 
\tau_{\rm{cut}}= 10, p=2$.
The upper plots are 3D maps 
in a box of $x \times y \times z = 
150\:\rm{km} \times 150\:\rm{km} \times 150\:\rm{km}$,
while the lower ones are projections 
on the x-y plane. 
We assume as threshold density
below which we identify the 
wind region a value of 
$5 \cdot 10^{9}\:\rm{g\:cm^{-3}}$.
However, this limit in the end 
also includes the outer regions
of the disk, visible as $\dot{e}<0$ 
regions around z=0 km.
The plots show the particle 
distribution, and we recover the 
values of $\dot{e}$ and $\dot{Y}_{\rm{e}}$
of each SPH particle from 
M1 and \Se via interpolation
from the respective grids.
Overall, both the ASL and the M1 
$\dot{e}$ and $\dot{Y}_{\rm{e}}$
distributions
agree well with the solution from \se.
However, they both
show some cooling ($\dot{e}<0$)
above the remnant which is weaker
or absent
in \se, as well as 
a few particles with $\dot{Y}_{\rm{e}}<0$.
More precisely, for the ASL 
$\approx 14\%$ of the particles have 
$\dot{e}_{\rm{ASL}} \cdot \dot{e}_{\rm{Sed}}<0$,
and $\approx 3\%$ have 
$\dot{Y}_{e,\rm{ASL}} \cdot \dot{Y}_{e,\rm{Sed}}<0$.
For the M1, these numbers are
$\approx 27\%$ and $\approx 1\%$
respectively.
Considering the overall  
similarity between
the maps of the ASL and M1 
we can conclude within the limits
of this analysis that the ASL may
show similar performances of M1 
in dynamical simulations when
comparing the wind properties 
against exact solutions to
the transport.
Nevertheless, a more robust
assessment requires to explore 
this comparison dynamically and
for different binary configurations.
The main advantage of our SPH-ASL 
is in its
efficiency. We 
indeed estimate that, 
if we had to assume the 
same timestep
as in M1 for a dynamical 
simulation,
and by taking MAGMA2 
as our Lagrangian hydrodynamics code
\citep{Rosswog2020},
the ratio between the 
CPU hours spent for the 
transport and those for the
hydrodynamics is about 0.8 per 
timestep, to be compared
with a factor of 
10 for the M1 in \fla.
The number we find is similar
to the one in \cite{pan19}.
The ratio for the SPH-ASL 
could be further cut down
if we consider that the
optical depth computation
may not be required
at every timestep,
unless the thermodynamical 
and compositional properties
of the matter change considerably.
Future dynamical simulation
will definitely provide a more robust 
assessment of the performance.

\section{Conclusions}
\label{sec:conclusions}
In this paper we have provided
a detailed calibration analysis of the
Advanced Spectral Leakage (ASL) scheme
presented earlier in \cite{Gizzi2019},
which is based on the 
original work of \cite{Perego2016}.
Our main motivation is the study of
neutrino-driven winds emerging from
neutron star mergers.
The gauging process is performed
by post-processing a number of 
snapshots of a binary neutron star 
merger remnant. We extract neutrino
quantities directly impacted
by each parameter,
and we compare the ASL results 
for different parameter combinations
with the ones 
obtained from the Monte Carlo neutrino 
transport code \Se \citep{Richers2015} and
from a two-moment scheme (M1) implemented
in \Fla \citep{Fryxell2000,OConnor2015,Oconnor2018}.
In the calibration process we focus on electron-type neutrinos and
anti-neutrinos, since
they determine the properties of neutrino-driven winds.
We summarise our main findings as follows:\\\\
$\bullet$ Performing neutrino 
transport in 2D by post-processing 
initial 3D post-merger
configurations is of limited accuracy at early 
($t \lesssim 20$ ms) times post-merger. 
In particular, the 2D averaging 
can severely
impact the total luminosities by more
than a factor of 2.\\\\
$\bullet$ The assumption of axially symmetric
neutrino fluxes entering the heating rate
in the ASL is validated by 3D neutrino 
transport simulations. In particular,
variations of the fluxes with 
the azimuthal angle $\phi$
and for a given 
polar angle $\theta$ are
limited in the bulk 
region of neutrino-driven winds.\\\\
$\bullet$ In agreement with 
\cite{Perego2016}, the 
thermalization parameter
$\tau_{\rm{cut}}$
has an impact mainly on the neutrino
mean energies. However, a value of 
$\tau_{\rm{cut}}= 10$ robustly
recovers neutrino mean energies
within $\lesssim 25\%$ accuracy.\\\\
$\bullet$ The heating parameter
introduced in \cite{Gizzi2019}
is re-calibrated to a lower
value, as a result of the
usage of \Se
as reference solution.
We find $p=2$ to best reproduce
the distribution of the neutrino
fluxes in all the cases examined.
Using a 
M1 scheme rather than 
\Se would lead to the 
artifact of a larger $p$ 
than the ones 
calibrated here
because of
the approximations introduced by the 
analytical closure.\\\\
$\bullet$ The blocking parameter $\alpha_{\rm{blk}}$
mainly impacts the total 
neutrino luminosities, in agreement with the results
of 
\cite{Perego2016}.
Moreover, unlike the other two
parameters it 
is the most sensitive to a 
change of binary configuration.\\\\
$\bullet$ The assumption 
$\alpha_{\rm{blk}} \equiv \alpha_{\nu_e,\rm{blk}}=
\alpha_{\bar{\nu}_e,\rm{blk}}$
adopted in \cite{Perego2016} 
can be rather inaccurate for
recovering neutrino luminosities.
In particular, 
electron neutrino luminosities
are lower by up 
to a factor of 2 with respect
to the reference solution,
suggesting
$\alpha_{\nu_e,\rm{blk}} <
\alpha_{\bar{\nu}_e,\rm{blk}}$.\\\\
$\bullet$ Assuming 
$\alpha_{\nu_e,\rm{blk}} \neq
\alpha_{\bar{\nu}_e,\rm{blk}}$,
we indeed find that 
$\alpha_{\nu_e,\rm{blk}}= 0.45$
provides electron neutrino
luminosities in agreement with 
the reference solution at the level
of $\lesssim 20\%$ accuracy, for both
equal and unequal mass binaries.
Similarly, $\alpha_{\bar{\nu}_e,\rm{blk}}= 0.55$
results in electron anti-neutrino
luminosities off by $\lesssim 20\%$
for equal mass binaries.\\\\
$\bullet$ We find 
$\alpha_{\bar{\nu}_e,\rm{blk}} = 0.75$
for unequal mass binaries, leading
to a relative error 
in the anti-neutrino
luminosity
of $< 30\%$. However, more 
test cases might be needed for 
a more robust evaluation and to
reduce systematics.\\\\
$\bullet$ In contrast to 
\cite{Foucart2020},
the heavy-lepton neutrino
luminosity is systematically larger by a factor
of a few in both the ASL and M1 with respect
to an exact solution to the transport. 
This enhances the overall cooling
of the remnant in our snapshot
calculations. For the ASL, the most 
probable explanation is the poor treatment of
the diffusion timescale,
which according to \cite{Ardevol19}
can boost luminosities by more 
than a factor of 2.
We expect this treatment to
affect also the electron neutrino
and anti-neutrino luminosities to
some extent.
\\\\
$\bullet$ The properties of neutrino-driven winds
are shaped by the rates of 
change of internal energy and electron 
fraction, $\dot{e}$ and $\dot{Y}_e$, respectively. 
The corresponding maps for a
1.4-1.4 $M_{\odot}$ binary demonstrate 
that for our suggested parameter 
choice the ASL scheme 
performs similar to the M1 approach.
From the perspective of dynamical
simulations, our SPH-ASL comes
with the advantage of
a better efficiency. 
In particular,
by taking the 
Lagrangian hydrodynamics 
code MAGMA2 \citep{Rosswog2020},
we estimate 
that the ratio between
the CPU hours 
spent for the transport
and those
spent for the hydrodynamics would be
$\lesssim 0.8$ per timestep, 
while for the 
M1 in \Fla is about 10. In other words, 
the ASL scheme could be applied in SPH simulations
with only a moderate additional computational effort.\\\\
Although the geometry of a binary 
neutron star merger allows neutrinos
to escape with more directional freedom
than in a spherically symmetric
core-collapse supernova, the results we
find here show that the blocking parameter
can still be quite high in merger remnants.
Apart from the different thermodynamics and
composition of the matter, 
the major reason is 
the disk geometry that
increases the effect
of inward neutrino fluxes
at the neutrino surfaces
with respect to a spherically
symmetric geometry.
In this paper we have also presented 
a completely mesh-free, particle-based algorithm
to compute spectral, species-dependent
optical depths, based on the Smoothed-Particle
Hydrodynamics (SPH) \citep{monaghan92,Monaghan05,Rosswog2009,Rosswog15,Rosswog2015_2,Rosswog2020}. 
This algorithm makes 
our ASL fully grid-independent, and
therefore suitable for future SPH dynamical 
simulations of binary neutron star mergers
with neutrino transport.
 
\section*{Acknowledgements}
We would like to thank Sherwood Richers for 
developing and supporting our use of 
\Se and for a careful reading of the manuscript.
This work has been supported by the Swedish Research 
Council (VR) under grant number 2016- 03657\_3, by 
the Swedish National Space Board under grant number 
Dnr. 107/16, by the Research Environment grant 
"Gravitational Radiation and Electromagnetic Astrophysical
Transients (GREAT)" funded by the Swedish Research 
council (VR) under Dnr 2016-06012 and by the Knut and Alice Wallenberg Foundation under Dnr KAW 2019.0112. EOC is supported by the Swedish Research Council (Project No. 2018-04575 and 2020-00452).
We gratefully 
acknowledge support from COST Action CA16104 
"Gravitational waves, black holes and fundamental 
physics" (GWverse), from COST Action CA16214 
"The multi-messenger physics and astrophysics of 
neutron stars" (PHAROS), and from COST Action MP1304 "Exploring 
fundamental physics with compact stars (NewCompStar)".\\
The simulations were performed on resources provided
by the Swedish National Infrastructure for Computing (SNIC) at PDC (Centre for High Performance Computing) and on
the facilities of the North-German Supercomputing Alliance
(HLRN) in G\"ottingen and Berlin.

\section*{Data Availability}
The data concerning the initial conditions needed
to run the simulations, as well as the extracted 
neutrino quantities from all neutrino transport 
codes and neutrino-driven wind maps
are available from the authors
of this manuscript upon request.


\bibliographystyle{mnras}
\bibliography{References.bib}





\bsp	
\label{lastpage}
\end{document}